\documentclass[journal,draftcls,onecolumn,12pt,oneside]{IEEEtran}

\usepackage[utf8]{inputenc}

\usepackage{amsmath,amssymb,amsfonts,amsthm,array}
\usepackage{mathrsfs}

\usepackage{stmaryrd}

\allowdisplaybreaks

\newtheorem{theorem}{{Theorem}}
\newtheorem{lemma}{{Lemma}}
\newtheorem{corollary}{{Corollary}}
\newtheorem{definition}{{Definition}}

\usepackage{algorithmic}
\usepackage{graphicx}
\usepackage{xcolor}
\usepackage{tikz}
\usepackage{cite}
\usepackage[english]{babel}
\usepackage{balance}
\usepackage[justification=centering]{caption}
\usepackage[shortlabels]{enumitem}
\usepackage{subcaption}

\newcommand{\set}[1]{\mathcal #1} 
\renewcommand{\vec}[1]{\boldsymbol{#1}} 
\newcommand{\mat}[1]{\boldsymbol{\underline{#1}}} 
\newcommand{\prob}[2]{\mathbb{P}_{#2}\left[#1\right]} 
\newcommand{\Exp}[2]{\mathbb{E}_{#2}\left[#1\right]} 
\newcommand{\Var}[2]{\mathbb{V}_{#2}\left[#1\right]} 
\newcommand{\Cov}[2]{\mathbb{C}_{#2}\left[#1\right]} 
\newcommand{\defeq}{\overset{\bigtriangleup}{=}}

\title{Distributed Source Coding for Parametric and Non-Parametric Regression}

\author{Jiahui Wei, \IEEEmembership{Student Member, IEEE}, Elsa Dupraz, \IEEEmembership{Member, IEEE},  Philippe Mary, \IEEEmembership{Member, IEEE} \thanks{J. Wei and P. Mary are with Univ Rennes, INSA Rennes, CNRS, IETR-UMR 6164, F-35000 Rennes, France.}\thanks{J. Wei and E. Dupraz are with IMT Atlantique, CNRS UMR 6285, Lab-STICC, Brest, France} \thanks{This work has received a French government support granted to the Cominlabs excellence laboratory and managed by the National Research Agency in the ``Investing for the Future'' program under reference ANR-10-LABX-07-01. This work was also funded by the Brittany region.} \thanks{Part of the content of this paper has been published in Eusipco 2023 and IZS 2024. We extend the analysis to  the non-trivial case of kernel methods and investigate the non-asymptotic trade-off between data reconstruction and regression.}
}

\begin{document}

\maketitle

\tikzset{
box/.style ={
rectangle, 
rounded corners =5pt, 
minimum width =50pt, 
minimum height =20pt, 
inner sep=5pt, 
draw=black },
bigbox/.style ={
rectangle, 
minimum width =50pt, 
minimum height =65pt, 
inner sep=5pt, 
draw=black 
},
roundnode/.style={
circle, draw=black, 
very thick, 
minimum size=6mm},
squarenode/.style={
rectangle, 
draw=black, 
very thick, 
minimum size=7mm}
}
 
\begin{abstract}
The design of communication systems dedicated to machine learning tasks is one key aspect of goal-oriented communications. In this framework, this article investigates the interplay between data reconstruction and learning from the same compressed observations, particularly focusing on the regression problem.
We establish achievable rate-generalization error regions for both parametric and non-parametric regression, where the generalization error measures the regression performance on previously unseen data.
The analysis covers both asymptotic and finite block-length regimes, providing fundamental results and practical insights for the design of coding schemes dedicated to regression. The asymptotic analysis relies on conventional Wyner-Ziv coding schemes which we extend to study the convergence of the generalization error. The finite-length analysis uses the notions of information density and dispersion with additional term for the generalization error. We further investigate the trade-off between reconstruction and regression in both asymptotic and non-asymptotic regimes. 
Contrary to the existing literature which focused on other learning tasks, our results state that in the case of regression, there is no trade-off between data reconstruction and regression in the asymptotic regime. We also observe the same absence of trade-off for the considered achievable scheme in the finite-length regime, by analyzing correlation between distortion and
generalization error. 
\end{abstract}

 
\section{Introduction}

\subsection{Context and problem}



The interaction of machine learning and communications is presently a vibrant area of research with numerous works dedicated to machine learning for communications. 
But the reciprocal relationship is also of significant interest and lies in the design of communication systems dedicated to machine learning tasks. This paradigm falls into the emerging area of goal-oriented communications~\cite{Shi2023}. 
In this case, the primary objective of the communication system shifts towards extracting and transmitting relevant information for the targeted learning task, encompassing methods such as hypothesis testing~\cite{ahlswede1986hypothesis}, regression~\cite{raginsky}, or classification~\cite{ehrlich2019deep}.
When engineers tackle machine learning over a rate-limited channel, the following key questions emerge: do the optimal encoder and decoder design for the learning task align with those used in conventional communication systems? Or is there an inherent trade-off between the learning task and data reconstruction? 
In this article, we address these questions in the context of regression.

Regression is one of the most popular supervised machine learning tasks and despite its apparent simplicity, it is used in many practical signal processing and telecommunication problems. For instance, non-parametric regression is used to reconstruct electrocardiograms in \cite{Onak2022}. In~\cite{Fabro2022}, the base station relies on the users signal feedback to estimate a radio map with a semi-parametric regression. In \cite{Angelosante2010}, frequency hopping parameters are inferred from a sparse linear regression. The work in \cite{Pawlak2007} deals with the identification of non-linearities in Wiener systems from a semi-parametric regression technique. In \cite{Wang2009}, a parametric regression technique is proposed to estimate a field function through distributed and noisy sensor measurements sent to a fusion center. 
However, none of these works considers the problem of compressing and communicating the data so as to perform the regression task at the remote server.

In this paper, we formulate the distributed regression problem as follows. As in standard regression, we consider a pair of real-valued random variables $(X,Y)$.  We assume that the statistical relation between $X$ and $Y$ is described by a function $f:\mathbb{R} \rightarrow \mathbb{R}$ such that $X=f(Y) + N$, where $f$ is unknown and $N$ is a Gaussian noise.  Like in the conventional Wyner-Ziv setup~\cite{wyner1976rate}, we consider that $X$ acts as the source to be encoded, while $Y$ serves as side information available only at the decoder. But in our case, the objective of the decoder is to infer the function $f$ from $Y$ and from the coded version of $X$. This regression is performed by minimizing the Mean-Squared Error (MSE) $\Exp{(\hat{f}(Y) - X)^2}{}$ with respect to $\hat{f}$.  

In cases where there is prior knowledge about the structure of the function $f$ (linear, polynomial, etc.) and $f$ depends on a finite number of parameters, the problem is termed as parametric regression. In this context, the ordinary least squares (OLS) estimator is known for providing the best unbiased estimator~\cite{rencher2008linear}.
On the other hand, non-parametric regression does not make any assumption about the structure of the underlying function $f$. In this case, various methods, such as kernel methods, K-Nearest Neighbors (KNN), or modeling involving a local or global averaging over the training set are applicable~\cite{wasserman_all_np}.
In this paper, we investigate both parametric regression and non-parametric kernel regression.

As learning performance criterion, we consider the regression generalization error, defined as the MSE evaluated on test samples different from the training samples. Our first objective is to provide achievable rates under constraints on the generalization error, for parametric and non-parametric regressions. Our second objective is to investigate the trade-off between reconstruction and regression, whenever the coding scheme is required to satisfy not only the constraint on the generalization error, but also another constraint on the distortion on $X$.  


\subsection{Related works}
Coding for computation has been a long-standing area of research, extending the Wyner-Ziv setup to cases where the decoder aims to compute a function of the source and side information.
Studies such as~\cite{koulgi2003zero,malak2020hyper}, building upon earlier works~\cite{orlitsky2001,yao1979}, have investigated the theoretical limits of coding data specifically for computation purposes. These studies typically focus on decoding one output value $f(X_k,Y_k)$ for each sample pair $(X_k,Y_k)$, using a predefined function $f$. However, this approach differs from our regression problem, where the goal is to infer the function $f$ itself across an entire length-$n$ sequence of sample pairs $\{(X_k,Y_k)\}_{k=1}^n$. Additionally, the theoretical frameworks in the previous works rely on the entropy of a characteristic graph, a measure suitable only for functions with finite support, rendering it less appropriate for addressing regression. 

Regarding coding schemes dedicated to learning,~\cite{Dobrushin1962, wolf1970} state that the optimal performance is achieved through an estimate-and-compress strategy. However, practical limitations due to hardware constraints or computational capabilities at the encoder, as well as the distributed nature of data across networks, often make this strategy impractical. In such cases, a compress-and-estimate scheme \cite{liu2021rate,Stavrou2022,el2015slepian,katz2017} is more relevant.
For example,  a rate-distortion framework has been introduced in~\cite{liu2021rate} specifically for semantic communications involving continuous sources, where each source is subject to its own distortion constraints: one for the information observed at the encoder and another for the hidden semantic source. 
This approach was further extended to discrete sources in~\cite{Stavrou2022}. 
Moreover, the information bottleneck framework~\cite{tishby2000information,tishby2015deep,pazone2022}  utilizes mutual information to measure the relevance of information extracted from the source.  Despite these advancements, there remains a significant challenge in linking distortion or mutual information metrics directly to the performance of the considered learning tasks. 

Some other works have considered coding with performance metrics specific to the considered learning task. 
Especially, the study conducted in~\cite{el2015slepian} demonstrated that, under the criterion of the variance of unbiased estimator, the rate necessary for estimating a parameter $\theta$ of the joint distribution $P_{XY}$ is lower than that required for source reconstruction.
Distributed hypothesis testing has also been widely explored recently. For instance,~\cite{katz2017,salehkalaibar2019,sreekumar2020} provided Type-II error exponents under constraints on the Type-I error for various hypothesis testing problems including testing against  independence. 
%
%
In addition, Raginsky has established lower and upper bounds on the learning generalization error of coding schemes dedicated to a range of distributed learning problems involving two sources $X$ and $Y$~\cite{raginsky}. However, we demonstrated in~\cite{jwei23, jwei23IZS} that the upper bound in~\cite{raginsky} is loose for both linear and polynomial regression. Building on these findings, this paper aims to investigate regression more broadly, addressing both parametric and non-parametric regression.


In the context of parameter estimation~\cite{el2015slepian}, hypothesis testing~\cite{katz2017}, as well as in the rate-distortion framework for semantic compression~\cite{liu2021rate, Stavrou2022, stavrou2023}, research has consistently demonstrated an inherent trade-off between data reconstruction and the specific learning task being considered. 
A similar trade-off was showcased for the problem of visual perception versus data reconstruction~\cite{blau2019rethinking}, where visual perception was measured from a divergence term, and also in the context of data identification and reconstruction in a noisy database~\cite{Tuncel2014}. 
In this paper, we also investigate this trade-off for the considered regression problems. 



\subsection{Contributions}
In this paper, we provide rate-generalization error regions for both parametric regression and kernel regression,  across both asymptotic and finite block-length regimes, for the source coding setup with side information.

We first investigate the asymptotically achievable rates under generalization error constraints. We utilize standard methods from asymptotic information theory,  \emph{e.g.},~\cite{wyner1976rate, draper2004universal}, which we extend to address the regression problem and analyze the generalization error. More specifically, we consider the achievable coding scheme of Draper~\cite{draper2004universal},
originally proposed for Wyner-Ziv coding when the joint distribution $P_{XY}$ is unknown.  
Within this scheme, our novel contribution lies in analyzing the convergence of the generalization error for regression, instead of  the distortion.
Our results demonstrate that the minimum expected regression generalization error can be achieved at any positive rate, thus closing the gap between the lower and upper bounds on the generalization error, and improving upon the upper bound established by Raginsky~\cite{raginsky}.

Furthermore, we extend these results beyond the asymptotic regime by employing finite block-length tools from~\cite{kostina,kostina_verdu,watanabe2015nonasymptotic}. 
Especially, while the original information density vector for the Wyner-Ziv problem comprised three components, two for the rate and one for the distortion~\cite{watanabe2015nonasymptotic}, we introduce an additional term accounting for the regression generalization error. This allows us to provide an achievable finite block-length rate-generalization error region for regression. 

Finally, in both the asymptotic and the finite block-length regime, we investigate the trade-off in terms of coding rate between regression and data reconstruction. Interestingly, our asymptotic analysis reveals a noteworthy outcome: contrary to findings in existing literature, there appears to be no trade-off between data reconstruction and regression in our investigated context. This result comes from the fact that asymptotically the generalization error upper bound matches the lower bound. At finite-length, we propose a novel method to analyze the trade-off by investigating the correlation between the distortion and generalization-error. We show that for the proposed achievability scheme, there is again no trade-off between the two constraints at finite-length. This  analysis of the correlation could be easily extended to  other achievability schemes which may be derived for the regression-reconstruction problem in the future.  



The remaining of the paper is organized as follows. Section~\ref{sec:prob} presents the source model and describes the considered regression methods. Section~\ref{sec:ge} defines the generalization error as well as the coding scheme for regression. Section~\ref{sec:asymptotic} and Section~\ref{sec:fbl} provide the asymptotic and non-asymptotic rate-generalization error regions, respectively. 
Section~\ref{sec:simuls} provides numerical results in non-asymptotic regime for several regression problems.


\section{Regression problem} 
In this section, after providing the notation we will use throughout the paper, we define the statistical model we consider for the sources $X$ and $Y$. We then present the two regression methods we investigate in this paper: parametric regression with OLS, and non-parametric regression from kernel methods. 
\label{sec:prob}
\subsection{Notation}

A random variable $X$ is denoted with a capital letter, while a realization of the random variable is denoted with lower-case letter $x$. 
Let $\Exp{X}{}$ and $\Var{X}{}$ be the mean and variance of the random variable $X$. 
Random vectors of length $n$ are denoted in bold, \emph{e.g}, $\vec{X} = \left[X_1, ..., X_n\right]^T$, and  
$\mathbb{E}[\vec{X}]$ and $\Cov{\vec{X}}{}$ are the mean vector and the covariance matrix of $\vec{X}$, respectively. 
Next, we use bold letters with underlines, e.g., $\mat{X}$ to denote matrices. 
When $\mat{X}$ is a squared matrix, we use Tr($\mat X$) to denote its trace, while $\lambda_{\max}(\mat X)$ and $\lambda_{\min}(\mat X)$  are  the maximum and minimum eigenvalues of $\mat{X}$,  respectively. We further denote $||\mat{X}||$ as the norm-2 of a matrix $\mat{X}$.
Sets are denoted with calligraphic fonts,  and if $f: \set{X} \rightarrow \set{Y}$ is a mapping then $\left|f\right|$ is the cardinality of $\set{Y}$. In addition, $\log(\cdot)$ denotes the base-2 logarithm. Moreover, the indicator function is defined as $\mathbf{1}\left[x\in \set{A}\right] = 1$ if $x\in \set{A}$ and $0$ otherwise.

Let us consider the measurable space $\left( \set{X} , \mathscr{B}\left(\set{X}\right) \right)$, where $\mathscr{B}\left(\set{X}\right)$ is the Borel $\sigma$-algebra on the set $\set{X}$. The probability measure $P_X$ over $\left( \set{X} , \mathscr{B}\left(\set{X}\right) \right)$ 
is the distribution of $X$. 
The notation $\prob{\cdot}{}$ is used for the probability of an event over the underlying probability space.  When $\set{X} = \mathbb{R}$ and the Radon-Nykodym derivative of $P_X$ with respect to the Lebesgue measure $\lambda$ exists, then it is denoted $p_{X}$ and is called the probability density function of the random variable $X$. When $\mathcal{X}$ is countable or finite and the Radon-Nykodym derivative with respect to the counting measure $\mu$ exists, $p_X$ is referred to as the the probability mass function of this random variable.

Being given a random vector $\vec{X}$, the probability measure $P_{\vec{X}}$ on the measurable space $\left(\mathbb{R}^n , \set{B}\left(\mathbb{R}^n\right)\right)$ admits a joint probability density function $p_{\vec{X}}$ if for all $\vec{x}=\left[x_1, x_2, \cdots, x_n\right]^T$ in $\mathbb{R}^n$ we have:
\begin{align}
    &P_{\vec{X}}\left(\left]-\infty, x_1\right] \times \cdots \times \left]-\infty , x_n\right]\right) 
    =
    \int_{-\infty}^{x_1} \cdots \int_{-\infty}^{x_n} p_{\vec{X}}\left(u_1,\cdots,u_n\right)d u_1 \cdots d u_n.
\end{align}

Moreover, for a joint probability measure $P_{XY}$ on $\set{X}\times \set{Y}$, the information density is denoted as~\cite{polyanskiy2010} 
\begin{equation}\label{eq:def_information_density}
    \iota\left(x,y\right) := \log \frac{d P_{Y\left|\right.X=x}}{d P_Y} \left(y\right),
\end{equation}
where the ratio above is the Radon-Nykodym derivative of the conditional measure $P_{Y\left|\right. X=x}$ with respect to the measure $P_Y$, in $y$. 
Given a pair $\left(X, Y\right)$ on the measurable space $\left(\mathbb{R}^2 , \mathscr{B}\left(\mathbb{R}^2 \right)\right)$ induced by the joint probability measure $P_{XY}$, then the function
\begin{equation}
    \begin{array}{ccc}
         \mathbb{R}^2 & \longrightarrow & \mathbb{R}^+\\
         (x,y) & \longmapsto & \frac{p_{XY}\left(x,y\right)}{p_{X}\left(x\right)}
    \end{array}
\end{equation}
is called the conditional probability density function of $Y$ given $X$ and is denoted $p_{Y\left|\right.X}\left(y\left|\right. x\right)$.

When $X$ is discrete and $Y$ is continuous, let us consider the measurable space $\left( \mathbb{N} \times \mathbb{R},  \set{P}\left(\mathbb{N}\right) \otimes \mathscr{B}\left(\mathbb{R}\right)\right)$, where $\set{P}\left(\mathbb{N}\right)$ is a partition of the set of integers. We define the joint probability measure $P_{XY}$ such as, for all $A\in \set{P}\left(\mathbb{N}\right)$ and $B\in \mathscr{B}\left(\mathbb{R}\right)$, we have
\begin{align}
    P_{XY}\left(A\times B\right) &\defeq \int_{A\times B} p_X (x )p_{Y|X}(y|x)d\mu(x) d\lambda(y)
    =\sum_{x\in A} p_X(x) \int_{B} p_{Y|X}(y|x)dy.
\end{align}
\subsection{Source definitions}
Let $(X,Y)\sim P_{XY}$ be a pair of jointly distributed real-valued random variables, where $X$ is the source to be encoded and $Y$ is the side information only available at the decoder. We assume that there exists a function $f:\mathbb{R} \rightarrow \mathbb{R}$ such that 
\begin{align}
\label{regression_model}
    X  = f(Y) + N,
\end{align}
where $N\sim \mathcal{N}(0, \sigma^2)$ follows a Gaussian distribution with mean $0$ and variance $\sigma^2$. We further suppose that $N$ is independent from $Y$. Without loss of generality but for simplicity, we consider $\Exp{Y}{} = 0$. 
We do not make any further assumption on the distribution of $Y$, except for kernel regression where the distribution support of source $Y$ has to be bounded. 
Therefore, our theoretical results apply to a wide range of distributions for $X$ and $Y$. In addition, the function $f$ between $X$ and $Y$ is deterministic, and we consider that it is unknown. The purpose of regression is to infer the function $f$ from a set of observations represented by $n$ independent and identically distributed (i.i.d.) sample pairs $\{(X_k, Y_k)\}_{k=1}^n$. There exists different types of regression, depending on what prior knowledge is available on the structure of the function~$f$.  


\subsection{Parametric
regression}
In the case of parametric regression, \eqref{regression_model} can be rewritten as~\cite{rencher2008linear} 
\begin{align}
\label{para_reg}
    X = \sum_{i=0}^{k-1} \beta_ih_i(Y) + N,
\end{align}
where $k$ is the order of the regression, and the functions $h_i:\mathbb{R}\rightarrow \mathbb{R}$ are fixed and known in advance, while the parameters $\beta_i$ are unknown, for all $i \in \llbracket 0,k-1\rrbracket $. Therefore, parametric regression reduces to estimating the parameter vector $\vec{\beta} = [\beta_0,\cdots,\beta_{k-1}]$. 

Here, we consider the OLS estimator, known for being the unbiased estimator with the minimal variance~\cite[Chapter 7]{rencher2008linear}. Let us define $\vec{Y_j^\star} = [h_0(Y_j), ..., h_{k-1}(Y_j)]^T$ $\in \mathbb{R}^k$, and $\mat{Y}^\star = [\vec{Y^\star_1}, ..., \vec{Y^\star_n}] \in \mathbb{R}^{k\times n}$. For given vectors  $\vec{X}$ and $\vec{Y}$, the OLS estimator $\hat{\vec{\beta}}$ is given by 
\begin{align}
\label{eq_ols}
    \vec{\hat{\beta}} = \left(\mat{Y}^\star{\mat{Y}^{\star}}^T\right)^{-1}\mat{Y}^\star\vec{X} .
\end{align}
According to the properties of OLS estimators, we have \cite[Chapter 7]{rencher2008linear}:
\begin{align}
\label{eq_ols_properties}
    \Exp{\vec{\hat{\beta}}}{} = \vec{\beta}\ \text{ and } \ \Cov{\vec{\hat{\beta}}|\vec{Y}}{} = \sigma_{X|Y}^2 \left(\mat{Y}^\star{\mat{Y}^\star}^T\right)^{-1},
\end{align}
where $\Cov{\vec{\hat{\beta}}|\vec{Y}}{}$ is the covariance matrix of $\vec{\hat{\beta}}$ given $\vec{Y}$ and $\sigma_{X|Y}^2$ is the conditional variance of $X$ given $Y$.

\subsection{Non-parametric regression}
In the case of non-parametric regression, no prior assumption on the form of the function $f$ is made, and we typically resort to various local or global smoothing techniques to estimate the regression function $ \Exp{X|Y=y}{}$~\cite{wasserman_all_np}.
In this paper, we consider the widely used kernel regression technique as an example, and we leave the extension to other non-parametric regression techniques for future works.

A one-dimensional kernel is any smooth, symmetric function $K:\mathbb{R} \rightarrow \mathbb{R}$ such that $\forall x \in  \mathbb{R}, K(x) \geq 0$, and the following relations hold~\cite{hardle2004nonparametric}
\begin{align}
    &\int_\mathbb{R} K(x)dx = 1, \quad \int_\mathbb{R} xK(x) dx = 0,\quad \text{and}  
    \quad 0\leq \int_\mathbb{R} x^2K(x)dx \leq \infty .
\end{align}
The Nadaraya-Watson Kernel regression over $(\vec{X},\vec{Y})$ is defined as \cite{wasserman_all_np}:
\begin{align}
\label{eq_kernel_regression}
    \hat{f}(Y) = \frac{\sum_{j=1}^n K\left(\frac{Y-Y_j}{h}\right)X_j}{\sum_{j=1}^n K\left(\frac{Y-Y_j}{h}\right)} .
\end{align} 
Here $h$ is a positive number referred to as the bandwidth. Essentially, $\hat{f}$ represents a local average of $\vec{Y}$ based on the kernel $K$. The choice of the kernel and of the parameter $h$ have been the subject of extensive research in statistical learning. It has been shown that estimators using different kernels have similar performance in terms of estimation loss $\Exp{(X - \hat{f}(Y))^2}{}$, while the choice of the bandwidth, which controls the smoothing, is of greater significance \cite{hastie2017generalized}. But our theoretical results are generic and will apply to different kernels~$K$ and to a range of values for $h$.  

\section{Coding scheme for regression}
\label{sec:ge}
In this section, we describe the coding setup we consider for regression, with one training phase and one inference phase. We then introduce the generalization error used to evaluate the regression performance and provide formal definitions of the considered coding scheme. 

\subsection{Training and inference phases}

\begin{figure}
    \centering
        \centering
        \begin{subfigure}{0.5\textwidth}
            \begin{tikzpicture}[every node/.style={font=\small}]
                \centering
                \node (X) at (0,-2) {$\vec{X}$};
                \node (beta) at (7.5,-2) {$\hat{f}^{(n)}(\mathbf{Z},.)$};
                \node[squarenode,line width=0.8pt] (encoder) at (1.5,-2) { Encoder};
                \node[squarenode,line width=0.8pt] (decoder) at (3.5,-2) { Decoder};
                \node[squarenode,line width=0.8pt] (est) at (5.5,-2) { Training};
                \node (Y) at (3.5, -3) {$\vec{Y}$};
                
                \draw[->,line width=0.6pt] (X) -- (encoder.west);
                \draw[->,line width=0.6pt] (encoder) -- (decoder.west);
                \draw[->,line width=0.6pt] (decoder) -- node[above] {$\vec{U}$} (est.west);
                \draw[->,line width=0.6pt] (est) -- (beta);
                \draw[->,line width=0.6pt] (Y) -- (decoder.south);
                \draw[->,rounded corners=3pt,line width=0.6pt] (Y.east) -| (5.5,-3) -- (est.south);
            \end{tikzpicture}
            \caption{Training phase}
        \end{subfigure}
        \centering
        \begin{subfigure}{0.5\textwidth}
            \begin{tikzpicture}[every node/.style={font=\small}]
                \node (Empty) at (0,-3) {};
                \node (Y2) at (3,-3) {$\tilde Y$};
                \node[squarenode,line width=0.8pt] (decoder2) at (3,-2) {Inference};
                \node (X2) at (6,-2) {$\hat{\tilde{X}} = \hat{f}^{(n)}(\mathbf{Z},\tilde Y)$};
                \draw[->,line width=0.6pt] (Y2) -- (decoder2.south);
                \draw[->,line width=0.6pt] (decoder2) -- (X2.west);
            \end{tikzpicture}
            \caption{Inference phase}
        \end{subfigure}
    \caption{ Coding scheme for regression, with one training phase (a) over the learning sequence $\mathbf{Z}=(\mathbf{U},\mathbf{Y})$ which provides a predictor $\hat{f}^{(n)}(\mathbf{Z},.)$, and one inference phase (b) which consists of applying the predictor on new samples~$\tilde{Y}$.} 
    \label{fig:test_channel_wz}
\end{figure}
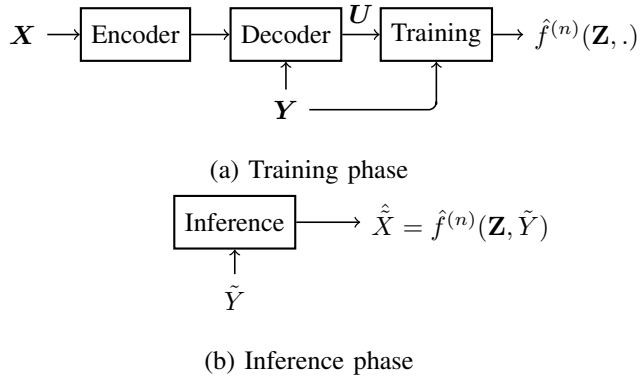

Regression, as a standard supervised learning problem, comprises one training phase and one inference phase, as shown in Figure \ref{fig:test_channel_wz}. 
A training sequence $\vec{Z} = (\vec{U},\vec{Y}) \in \mathcal{Z}^n$ of length $n$ is built using the available side information $\vec{Y}$ and a coded representation of $\vec{X}$, denoted as $\vec{U}$. 
The training phase aims to
estimate the function $f$ on $\mathbf{Z}$ from either parametric or non-parametric regression.
It provides a sequence of functions, called predictors, denoted as $\hat{f}^{(n)}:\mathcal{Z}^n \times \mathbb{R} \rightarrow \mathbb{R}$. 
It is important to mention that the training sequence involves the coded sequence $\vec{U}$ because the decoder does not have direct access to $\vec{X}$. Therefore, \eqref{eq_ols} and~\eqref{eq_kernel_regression} need to be updated so as to account for $\vec{U}$, as will be described in Section~\ref{sec:asymptotic}. 

Next, we use $\tilde{X}$ and $\tilde{Y}$ to denote random variables from the inference phase, where the pair $(\tilde{X}, \tilde{Y})$  follows the same probability distribution  $P_{XY}$ of the pair $(X,Y)$ while being independent from it. At the inference phase, the decoder uses the predictor $\hat{f}^{(n)}$ to produce estimates of $\tilde{X}$ as $\hat{\tilde{X}} = \hat{f}^{(n)}(\mathbf{Z},\tilde Y) $. 
It is worth noting that this does not require any data transmission, since the side information $\tilde{Y}$ is already available to the decoder. Therefore, this paper investigates the coding scheme for the training phase only, while the performance of this scheme is evaluated over the inference phase with the generalization error. 



\subsection{Generalization error}
Usually, the performance of a lossy source coding scheme is evaluated from a distortion measure. 
Since here the objective of the receiver is also to learn a regression function, we need to consider additional metrics relevant for the regression problem. For that purpose, we use the notions of expected loss and generalization error already considered in~\cite{raginsky}. 

A quadratic loss function $\ell: \mathbb{R}^2 \rightarrow \mathbb{R}$ defined as $\ell(x,\hat{x}) = (x - \hat{x})^2$ is considered. The expected loss $L$ is defined as:
\begin{equation}
L(f) = \Exp{\ell(X,f(Y))}{}.
\end{equation}
For a given regression problem, let $\mathcal{F}$ represents the set of regression functions of the form $f : \mathbb{R} \rightarrow \mathbb{R}$, with a predefined form (parametric regression) or free of specific assumptions (non-parametric regression). For instance, for polynomial regression, $\mathcal{F}$ is the set of all polynomial functions of a fixed order $k$.  The minimum expected loss $L^{\star}$ is then given by:
\begin{equation}
\label{minimum_expected_loss}
L^{\star}(\mathcal{F}) = \inf_{f \in \mathcal{F}} L(f).
\end{equation}
Next, the generalization error is defined as:
\begin{equation}
\label{eq_generalization_error}
G(\hat{f}^{(n)}, \vec{Z}) = \Exp{\ell\left( \tilde{X},\hat{f}^{(n)}(\vec{Z},\tilde{Y}) \right) \left|\right. \vec{Z}}{\tilde X \tilde Y}.
\end{equation}
The generalization error defined in~\eqref{eq_generalization_error} is a random variable due to the conditioning on $\vec{Z}$. Therefore, in what follows, we will also resort to the expected generalization error $\Exp{G\left( \hat f^{(n)},\vec{Z} \right)}{\vec{Z}}$.

The minimum expected loss defined in~\eqref{minimum_expected_loss} is reached for the function $f^{\star}$ that minimizes the quantity $\Exp{\ell(X,f^{\star}(Y))}{}$ over the space of functions $\mathcal{F}$. However, there is no guarantee that this optimal function $f^{\star}$ can be  estimated exactly from the training sequence $\mathbf{Z}$.
Therefore, the generalization error measures the average quadratic loss which can be achieved for a specific training sequence $\vec{Z}$ and for a given predictor $\hat{f}^{(n)}(\vec{Z}, \cdot) $. The generalization error is evaluated as the expectation over the distribution $P_{\Tilde{X}\Tilde{Y}}$ of the MSE between the symbol $\tilde X$ and the estimated symbol $\hat{\tilde X} =\hat{f}^{(n)}(\vec{Z}, \tilde{Y})$. In our case, we assume that $(\tilde{X},\tilde{Y})$ follows the same distribution as $(X,Y)$, but \eqref{eq_generalization_error} would also apply otherwise. 

In addition, by bias-variance decomposition, it can be shown that the expected generalization error $\Exp{G\left( \hat f^{(n)}, \vec{Z}\right)}{\vec{Z}}$ is lower bounded as
\begin{equation}
L^{\star}(\mathcal{F}) \leq \Exp{G\left( \hat f^{(n)}, \vec{Z}\right)}{\vec{Z}} .
\end{equation}
Therefore, the difference $\delta = \Exp{G\left( \hat f^{(n)} ,\vec{Z}\right)}{\vec{Z}} - L^{\star}(\mathcal{F})$ is a crucial quantity for characterizing the performance of a regression coding scheme. This is why our rate-generalization error regions defined in the next section will be expressed with this quantity.

\subsection{Coding scheme}
In~\cite{jwei23}, we introduced a coding scheme which was initially dedicated to the linear regression problem, by adapting definitions from~\cite{raginsky}. But this scheme is actually generic enough to be adopted for any type of parametric regression, and for non-parametric regression as well.  This is why we restate it here. 
\begin{definition}\label{def:reg_scheme}
A regression scheme at rate $R$ is defined by a sequence $\{(e_n, d_n, R, t_n)\}$ with an encoder  $e_n: \mathcal{X}^n \xrightarrow[]{} \llbracket 1,M_n\rrbracket$,
a decoder $d_n : \mathcal{Y}^n \times \llbracket 1,M_n\rrbracket \rightarrow \mathcal{U}^n$,
and a learner $t_n: \set{Y}^n \times \set{U}^n \rightarrow \mathcal{F}$, 
such that
\begin{equation}
    \notag
    \mathop{\lim\sup}\limits_{n\rightarrow\infty} \frac{\log M_n}{n} \leq R .
\end{equation}    
\end{definition}
\begin{definition}\label{def:nml}
    An $(n, M, G)$ code for the sequence $\{(e_n, d_n, R, t_n)\}$ is a code with $|e_n| = M_n$ such that 
    \begin{equation}
    \Exp{G(\hat{f}^{(n)}, \vec{Z})}{} \leq G\  \text{and} \ \mathop{\lim\sup}\limits_{n\rightarrow\infty} \frac{\log M_n}{n} \leq R .
    \end{equation}
\end{definition}
\begin{definition} 
\label{def:r_delta}
     A pair $(R,\delta)$ is said to be achievable if an $(n,M,G)-$code exists such that
\begin{equation}
    \limsup_{n\rightarrow\infty} \Exp{G(\hat{f}^{(n)}, \vec{Z})}{\vec{Z}} \leq  L^*(\mathcal{F}) +  \delta .
\end{equation}
\end{definition}
As discussed in the previous section and similar to the definition used in \cite{raginsky}, the achievable region is defined in terms of the gap between $\Exp{G(\hat{f}^{(n)})}{\vec{Z}}$ and $L^*(\mathcal{F})$.

In this paper, we also consider the case where the decoder may either want to reconstruct the source $X$, or perform regression. In this case, the reconstruction task is evaluated with the standard quadratic distortion measure $d(x, \hat{x}) = (x - \hat{x})^2$, where $\hat{x}$ is the reconstruction of $x$ at the decoder. 
We further define the coding scheme with both reconstruction and regression constraints as follows. 
\begin{definition}\label{def:nmdl}
    An $(n, M, D, G)$ code for the sequence $\{(e_n, d_n, R, t_n)\}$ is a code with $|e_n| = M$ such that 
    \begin{equation}
     \Exp{d(X,\hat X)}{} \leq D \ , \ \Exp{G(\hat{f}^{(n)}, \vec{Z})}{} \leq G \ , \    \text{and} \ \frac{\log M_n}{n} \leq R .
    \end{equation}
\end{definition}  

We now provide definitions for the finite-length analysis of the coding schemes. 
\begin{definition}\label{def:nmleps}
    An $(n, M, G, \varepsilon)$ code for the sequence $\{(e_n, d_n, R, t_n)\}$ and $\varepsilon \in (0,1)$ is a code with $|e_n| = M_n$ such that 
    \begin{equation}
    \prob{G(\hat{f}^{(n)}, \vec{Z}) \geq G}{ }{} \leq \varepsilon\  \text{and} \ \frac{\log M}{n} \leq R .
    \end{equation}
\end{definition}

\begin{definition}
    For fixed $G$ and block-length $n$, the finite block-length rate-loss function with excess loss $\varepsilon$ is defined by:
    \begin{equation}
    \begin{aligned}
        R(n, G, \varepsilon) &= \inf_{R} \{\exists \ \  (n, M, G, \varepsilon) \ code \} .
    \end{aligned}
    \end{equation}
\end{definition}

\begin{definition}\label{def:nmdleps}
    An $(n, M, D, G, \varepsilon)$ code for the sequence $\{(e_n, d_n, R, t_n)\}$ and $\varepsilon \in (0,1)$ is a code with $|e_n| = M$ such that 
    \begin{subequations}
    \begin{align}
     \prob{\left\{d(X,\hat X) \geq D\right\} \ \cup \ \left\{G(\hat{f}^{(n)}, \vec{Z}) \geq G\right\}}{ } & \leq \varepsilon \\
     \frac{\log M}{n} & \leq R .
    \end{align}
    \end{subequations}
\end{definition}

\begin{definition}\label{def:reg_finitebl}
    For fixed $D$, $G$ and block-length $n$, the finite block-length rate-distortion-generalization error functions with excess loss $\varepsilon$ is defined by:
    \begin{equation}
    \begin{aligned}
        R(n, D, G, \varepsilon) &= \inf_{R} \{\exists \ \  (n, M, D, G, \varepsilon) \ code \} .
    \end{aligned}
    \end{equation}
\end{definition}

Definitions~\ref{def:reg_scheme} to~\ref{def:r_delta} will be used for the asymptotic analysis of Section~\ref{sec:asymptotic}, and are similar to what was initially introduced in~\cite{raginsky} and later considered in~\cite{jwei23}. On the other hand, Definitions~\ref{def:nmdl} to~\ref{def:reg_finitebl} did not appear in~\cite{raginsky}, and will serve to investigate the trade-off between data reconstruction and regression, as well as for the finite block-length analysis in Section~\ref{sec:fbl}.

\section{Asymptotic Analysis} \label{sec:asymptotic}

In~\cite[Theorem 3.3]{raginsky}, it is shown that, when considering a quadratic loss function, the expected generalization error can be lower and upper bounded as follows:
\begin{equation}
\label{eq:bounds_gen}
\begin{split}
L^{\star}(\mathcal{F})^\frac{1}{2} \leq \limsup_{n\rightarrow\infty}\Exp{G(\hat{f}^{(n)}, \vec{Z})^{\frac{1}{2}}}{} &\leq L^{\star}(\mathcal{F})^\frac{1}{2} 
+ 2\mathbb{D}_{X|Y}(R)^{1/2},
\end{split}
\end{equation}
where $\mathbb{D}_{X|Y}(R)$ represents the conditional distortion-rate function\cite{gray1972conditional}. It is worth noting that for regression problems considering the quadratic loss function, we have
\begin{align}
    L(\hat{f}) = \sigma^2 + \Exp{\left(\hat{f}(Y) - f(Y)\right)^2}{}\geq\sigma^2,
\end{align}
with equality iff $\hat{f} = f$. So the minimum expected loss defined in \eqref{minimum_expected_loss} is given by $L^{\star}(\mathcal{F}) = \sigma^2$.  
In this section, we propose a coding scheme which improves over the upper bound in~\eqref{eq:bounds_gen}, both for parametric and non-parametric regression.

\subsection{Achievable rate-generalization error regions}
The next two Theorems provide the rate-generalization error regions which can achieved for both parametric regression and kernel regression. 
\label{sec_asym_mr}
\begin{theorem}[Parametric regression]
\label{thm_parametric}
    Given any rate $R > 0$, the pair $(R, 0)$ is achievable for parametric regression with quadratic loss, for sources $(X,Y)$ following the model in \eqref{para_reg}.
\end{theorem}

This Theorem generalizes results we obtained in~\cite{jwei23,jwei23IZS} (linear and polynomial regression), to any type of parametric regression described by~\eqref{para_reg}.
It states that the minimum generalization error $L^{\star}(\mathcal{F})$ can be achieved with arbitrary rate $R$, as long as the length $n$ of the training sequence is large enough. 
Therefore, Theorem~\ref{thm_parametric} improves over the result of~\cite{raginsky} by eliminating the term $\mathbb{D}_{X|Y}(R)$ in the upper bound in~\eqref{eq:bounds_gen}. This makes our result tight in the sense that the upper bound equates the lower bound $L^{\star}(\mathcal{F})$ for any rate $R>0$. 


\begin{theorem}[Kernel regression]
\label{thm_kernel}
    Under the following conditions:  
    \begin{enumerate}[i.]
        \item $Y$ is bounded almost surely 
        \item the probability density function $p_Y$ is continuously differentiable and positively lower bounded, 
        \item the regression function $f$ is twice continuously differentiable, i.e. $f'$, and $f''$ exist,
        \item $h = h_n$ is a deterministic sequence such that when $n\rightarrow \infty$, the bandwidth $h$ satisfies $h\rightarrow 0$ and $nh\rightarrow \infty$,
    \end{enumerate}
    given any rate $R>0$, the pair $(R, 0)$ is asymptotically achievable for the kernel regression with quadratic loss.
\end{theorem}

Like for parametric regression, the previous Theorem states that kernel regression over the pair $(\vec{U},\vec{Y})$ can asymptotically achieve the same performance as when applied on original data $(\vec{X},\vec{Y})$. In fact, in the case of kernel regression, we will show that the generalization error can be divided into three parts: a first part for the intrinsic noise $N$ given by the term $\sigma^2$, a second and third part related to the bias and the variance of the estimator.
We show that the last two terms go to $0$ as $n$ goes to infinity because of condition iv in Theorem~\ref{thm_kernel}, in particular. In addition, the proof for the rate-generalization error region for kernel regression differs from the case of parametric regression, given that in the later case no prior assumption on the regression function is considered. But the conclusion is still that the gap $\Exp{G\left( \hat f^{(n)}, \vec{Z}\right)}{\vec{Z}} - L^{\star}(\mathcal{F})$ tends to $0$ as $n$ goes to infinity. We leave for future works the investigation of other methods, like local polynomial regression, which could further reduce the bias for finite $n$.

\subsection{Proof of Theorem~\ref{thm_parametric} and Theorem~\ref{thm_kernel}}
\label{subsec_proof}
We now briefly describe the achievability scheme that is considered in the proofs of Theorem~\ref{thm_parametric} and Theorem~\ref{thm_kernel}. We then provide expressions as well as convergence analysis of the generalization error for both parametric and kernel regression, since those constitute our technical contribution for the asymptotic case. 

In our considered achievability scheme, we make use of a Gaussian test channel described by $U = \alpha(X + \Phi)$, where $\Phi \sim \mathcal{N}(0, \sigma_\Phi^2)$ is independent of $X$. The parameters $\alpha$ and $\sigma_\Phi$ are constant and depend on the distributions of $X$ and $Y$. We then consider the achievability scheme proposed by Draper in~\cite{draper2004universal} for Wyner-Ziv coding in the case where the distribution $P_{XY}$ is unknown. This scheme is based on quantization and binning, and provides a criterion on empirical information density for debinning. We also use this criterion, but do not consider the incremental coding strategy of~\cite{draper2004universal} which is not necessary here as the coding rate is fixed given that $\sigma^2$ is known. This scheme is described into details in Appendix~\ref{apx:test_channel}. The results of~\cite{draper2004universal} show that the sequence $\vec{U}$ can be reconstructed by the decoder with vanishing error probability as $n$ tends to infinity.
We next demonstrate that the Gaussian test channel allows us to achieve the optimal rate-generalization error region for both parametric regression and kernel regression, by expressing the generalization error in both cases. 

\subsubsection{Convergence analysis of the generalization error in Theorem~\ref{thm_parametric}}
    For the parametric regression model described in~\eqref{para_reg}, the OLS estimator applied over the pair $(\vec{U}, \vec{Y})$ is given by
\begin{align}\label{eq:exp_beta_uy}
    \vec{\hat{\beta}} = \alpha^{-1}(\mat{Y}^\star{\mat{Y}^{\star}}^T)^{-1}\mat{Y}^\star\vec{U},
\end{align}
and it has the following properties:
\begin{align}
    \Exp{\vec{\hat{\beta}}}{} = \vec{\beta}\ \text{ and } \ \Cov{\vec{\hat{\beta}}|\vec{Y}}{} = \frac{1}{\alpha^2}\sigma_{U|Y}^2 (\mat{Y}^\star{\mat{Y}^\star}^T)^{-1}.
\end{align}
Note that this differs from what was defined in \eqref{eq_ols} and \eqref{eq_ols_properties} since the decoder has no direct access to $\vec{X}$.
From~\eqref{eq_generalization_error} and~\eqref{eq:exp_beta_uy}, the generalization error can be expressed as 
\begin{equation}
\label{eq_generalization_parametric}
    \begin{aligned}
        G(\hat{f}^{(n)}, \vec{Z}) 
        &= \Exp{[\vec{\beta} - \vec{\hat{\beta}} ]^T\vec{\tilde Y^\star} \vec{\tilde Y^\star}^T[\vec{\beta} - \hat{\vec{\beta}}] + N^2|\vec Z}{\tilde X\tilde Y} \\
        &=  [\vec{\beta} - \vec{\hat{\beta}} ]^T\Exp{\vec{\tilde Y^\star} \vec{\tilde Y^\star}^T}{\tilde Y}[\vec{\beta} - \vec{\hat{\beta}} ] + \sigma^2 ,
    \end{aligned}
\end{equation}
where $\vec{\tilde Y^\star} = [h_0(\Tilde{Y}), ..., h_{k-1}(\Tilde{Y})]^T$ 
refers to the vector composed of $h_i(\Tilde{Y}), \forall i\in  \llbracket 0,k-1\rrbracket$ independent from $Y$. 
By defining $\mat{\tilde{\Sigma}} = \Exp{\vec{\tilde Y^\star} \vec{\tilde Y^\star}^T}{\tilde Y}$ and $\mat{\Sigma} = \frac{1}{n}\mat{Y}^\star{\mat{Y}^\star}^T$, the expected generalization error can be expressed as 
\begin{align}
\notag
    &\Exp{G(\hat{f}^{(n)}, \vec{Z})}{\vec Z} \\ \notag
    &= \sigma^2 + \Exp{\frac{1}{n}(\mat{\Sigma}^{-1}\mat{Y}^\star(\vec N+\vec\Phi))^T \mat{\tilde{\Sigma}}\frac{1}{n}(\mat{\Sigma}^{-1}\mat{Y}^\star(\vec N+\vec\Phi))}{}\\
    &=\sigma^2 + \frac{\sigma^2 + \sigma_\Phi^2}{n}\Exp{\text{Tr}\left(\mat{\Tilde{\Sigma}}\mat{\Sigma}^{-1}\right)}{}\label{expectation_g_e} \\
    &\leq \sigma^2 + \frac{\sigma^2 + \sigma_\Phi^2}{n} \Exp{\frac{k\lambda_{\max}(\mat{\tilde{\Sigma}})}{\lambda_{\min}(\mat{\tilde \Sigma}) - ||\mat{\tilde \Sigma} - \mat{ \Sigma}||}}{}\\
    &\leq \sigma^2 + \frac{\sigma^2 + \sigma_\Phi^2}{n} \Exp{\frac{k\lambda_{\max}(\mat{\tilde{\Sigma}})}{\lambda_{\min}(\mat{\tilde \Sigma})}}{}\\
    &\leq \sigma^2 + \frac{\sigma^2 + \sigma_\Phi^2}{n} kC ,\label{eq:last_ineq_parametric}
\end{align}
where $C = \frac{\lambda_{max}(\mat{\Tilde{\Sigma}})}{\lambda_{min}(\mat{\Tilde{\Sigma}})}$ is a constant. 
When $n\rightarrow\infty$, the generalization error $\Exp{G(\hat{f}^{(n)}, \vec{Z})}{\vec Z}$ converges to $\sigma^2$, which completes the convergence analysis of Theorem \ref{thm_parametric}.

\subsubsection{Convergence analysis of the generalization error in Theorem~\ref{thm_kernel}}
Given the fact that the kernel regression is applied on the pair $(\vec{U},\vec{Y})$, and according to the Gaussian test channel $U = \alpha(X + \Phi)$, \eqref{eq_kernel_regression} can be rewritten as:
\begin{align}
\label{eq_kernel_uy}
    \hat{f}(y) = \frac{\sum_{i=1}^n K(\frac{y-y_i}{h})\frac{u_i}{\alpha}}{\sum_{i=1}^n K(\frac{y-y_i}{h})} .
\end{align}  

We now provide the main key steps of the convergence analysis of the generalization error. The details of the derivation are provided in Appendix~\ref{apx:kernel}.
For a given pair $( \tilde x,  \tilde y)$, the so-called test error can be expressed as: 
\begin{align}
\label{b-v_decomposition}
    \Exp{(\hat{f}^{(n)}( \tilde y, \vec{Z}) - f(\tilde y))^2}{\vec{Z}} =b_n^2(\tilde y) + V_n(\tilde y),
\end{align}
where $b_n(\tilde y) = \Exp{\hat{f}^{(n)}(\tilde y, \vec{Z}) - f(\tilde y)}{}$ is the bias and $V_n(\tilde y) = \Var{\hat{f}^{(n)}(\tilde y, \vec{Z})}{}$ is the variance of the estimator $\hat{f}^{(n)}$ with respect to the training sequence $\mathbf{Z}$. The expected generalization error is then given by 
\begin{align}
\label{eq_exp_ge}
    \Exp{G(\hat{f}^{(n)}, \vec{Z})}{\vec Z} &= \Exp{(\hat{f}^{(n)}( \tilde Y, \vec{Z}) - \tilde X)^2}{\tilde{X}\tilde{Y}\vec{Z}} \\
    &=  \sigma^2 \!\! + \!\!\! \int b_n^2(\tilde y) p_Y(\tilde y)d\tilde y 
    + \!\!\! \int V_n(\tilde y)p_Y(\tilde y) d\tilde y.
\end{align}
For a given $\Tilde{y}$, by analyzing the convergence of the numerator and the denominator of $\hat{f}(y)$ in \eqref{eq_kernel_uy}, it is shown in Appendix~\ref{apx:kernel} that 
\begin{align}
 \label{eq_bn}
    &b_n(\tilde y) = \frac{h^2}{2}\left(2\frac{f'(\tilde y)p'_Y(\tilde y)}{p_Y(\tilde y)} + f''(\tilde y)\right)\int_\mathbb{R} u^2K(u)du + o(h^2), \\ \label{eq_vn}
    &V_n(\tilde y) = \frac{(\sigma^2 + \sigma_\Phi^2)}{p_Y(\tilde y) nh}\int_\mathbb{R} K^2(u)du + o\left(\frac{1}{nh}\right).
\end{align}
Finally, as $n\rightarrow\infty, h\rightarrow 0$ and $nh\rightarrow 0$ (by condition iv in Theorem \ref{thm_kernel}), $\Exp{G(\hat{f}^{(n)}, \vec{Z})}{\vec Z}$ in \eqref{eq_exp_ge} tends to $\sigma^2$.


\subsection{Comparison with existing works}

First, we point out that authors in~\cite{kipnis2019} also proposed a Gaussian approximation of the quantization error under which the MSE of an estimator (not dedicated to regression) applied to compressed data is equivalent to the same estimator when applied to a corrupted version of data by a Gaussian noise. This is in line with our achievable scheme.

We now comment on our improvement of the upper bound of Raginsky in~\cite{raginsky}.  
First of all, the results of~\cite{raginsky} are stated for a generic loss function $\ell$ which can then be specified for various learning problems including regression or classification. In~\cite{raginsky}, the empirical loss  $\hat{L}_{\vec{X},\vec{Y}}(f)$ for a certain function $f$ is defined as
\begin{equation}
\hat{L}_{\vec{X},\vec{Y}}(f) = \frac{1}{n} \sum_{i=1}^n \ell(f(X_i),Y_i) ,
\end{equation}
and the difference between $\hat{L}_{\vec{X},\vec{Y}}(f)$ and $\hat{L}_{\vec{U},\vec{Y}}(f)$ is upper bounded as
\begin{equation}\label{eq:bounds_empirical_loss}
\sup_f |\hat{L}_{\vec{X},\vec{Y}}(f) - \hat{L}_{\vec{U},\vec{Y}}(f) | \leq \eta (d(\vec{U},\vec{X})),
\end{equation}
where $d$ is a distortion measure and $\eta$ is a concave
continuous function. Taking the expectation of~\eqref{eq:bounds_empirical_loss} as well as further mathematical manipulation lead to the upper bound on the generalization error in~\eqref{eq:bounds_gen}, especially given that $\mathbb{E}[d(\vec{U},\vec{X})] = \mathbb{D}_{X|Y}$. 
But in our case, expressing for instance the generalization error for parametric regression with quadratic loss and with the OLS estimator defined in~\eqref{eq:exp_beta_uy} gives that the term $\Exp{d(\vec{U},\vec{X})}{}$ (which is $\sigma_{\Phi}^2$ in~\eqref{eq:last_ineq_parametric}) is multiplied by a factor $1/n$ and therefore vanishes as $n$ goes to infinity. This is why we get that the generalization error converges to the minimum expected loss $\sigma^2$. As a result, the upper bound of~\cite{raginsky} is not tight in our setup, but on the other hand it applies to a larger range of learning problems. 

In addition, consider an alternative regression problem that is to infer a function $g$ such that $U = \alpha(g(Y) + N) + \Phi$, with $\Phi \sim \mathcal{N}(0,\mathbb{D}_{X|Y})$. For this alternative problem, the minimum expected loss~\eqref{minimum_expected_loss} expressed for $L(g) = \mathbb{E}[\ell(g(U),Y)]$ would be given by $\sigma^2 + \mathbb{D}_{X|Y}$, and hence we would retrieve the upper bound of~\eqref{eq:bounds_gen}. But here, since the target is to estimate the function $f$ such that $Y = f(X)+N$, it turns out that the minimum expected loss $\sigma^2$ can be achieved, despite  applying regression on the pair $(\vec{U},\vec{Y})$. 

\subsection{Regression-reconstruction trade-off}
\label{subsec:asymptotic_tradeoff}
Consider the achievability scheme described in Section \ref{subsec_proof}
for conventional Wyner-Ziv coding for reconstruction. This scheme achieves the Wyner-Ziv rate-distortion function provided in~\cite{wyner1978}, that is $R_{WZ}(D) = \inf I({{X}};{{U}}|{{Y}})$ where the $\inf$ is taken over $p_{U\left|\right.X}(u|x)$ and is such that the Markov chain $X\leftrightarrow U \leftrightarrow Y$ holds.  Given that we consider this same achievability scheme in our proofs of Theorem~\ref{thm_parametric} and Theorem~\ref{thm_kernel}, we can formulate a rate-distortion-generalization error function
as follows:
\begin{equation}
R\left( {{D},G} \right) = \mathop {\inf }\limits_{
\begin{array}{l} \qquad p(u\mid x) : \\ \Exp{ {d(X,\hat X)} }{} \leq {D} \\ \Exp{ G(\hat{f}^{(n)}, \vec{Z}) }{} \leq G 
\end{array}} 
 I({{X}};{{U}}|{{Y}}) .
\end{equation}

The next Corollary investigates the trade-off in $R(D,G)$ between the two constraints $ \Exp{ {d(X,\hat X)} }{} \leq {D}$ and $\Exp{ G(\hat{f}^{(n)}, \vec{Z}) }{} \leq G $. 


\begin{corollary}[Asymptotic trade-off for parametric and kernel regression]
\label{corollary_asymptotic}
    For a pair of sources $(X,Y)$ modeled from~\eqref{regression_model}, and for some non-negative constants $D$ and $G\geq \sigma^2$, we have
    \begin{equation} 
        R(D, G) = R_{WZ}(D)    
    \end{equation}
    for both parametric and kernel regression.
\end{corollary}
\begin{proof}
    See Appendix \ref{apx:corollary_1}.
\end{proof}


This results shows that the previous achievability scheme which minimizes the generalization error for regression can also achieve the optimal Wyner-Ziv rate-distortion function for reconstruction. Therefore, asymptotically there is no trade-off in terms of coding rate between reconstruction and regression. 
%
%
Note that this result differs from existing ones in the literature, which show that there is a trade-off between data reconstruction and other specific tasks, such as in the distortion-perception problem~\cite{blau2019rethinking}, for semantic communications~\cite{liu2021rate, Stavrou2022, stavrou2023}, for parameter estimation~\cite{ el2015slepian}, and for hypothesis testing \cite{katz2017}. 

The next section provides finite block-length rate-generalization error regions, and also investigates the trade-off between regression and reconstruction at finite length. 

\section{Finite block-length analysis} \label{sec:fbl}

The non-asymptotic source coding  problem with a distortion constraint and without side information was first investigated in~\cite{kostina, dispersion} using the notions of information density and dispersion region. This non-asymptotic analysis was also extended to the case with side information at the decoder in~\cite{watanabe2015nonasymptotic}. 
The main idea behind these analysis is to approximate the distribution of error events by the Berry-Esséen Theorem and to bound the resulting approximation error.  In this section, we extend these tools so as to also treat the regression problem. Especially, in finite block-length analysis, the excess probability $\epsilon$ which appears in Definition \ref{def:nmleps} plays a crucial role as not all the codewords satisfy the generalization error constraint.

In what follows, we directly address the source coding problem with the two objectives of data reconstruction and regression, and investigate the trade-off between these two tasks. The proposed analysis applies to both parametric and kernel regression.

\subsection{Definitions}

Let us consider the following four sets: 
\begin{align} 
\label{eq_tp}
\mathcal {T}_{ \mathrm {p}}(\gamma _{ \mathrm {p}}):=&\left \{{ (u,y) : \iota\left(u,y\right) \ge \gamma _{ \mathrm {p}} }\right \}, \\ \label{eq_tc}
\mathcal {T}_{ \mathrm {c}}(\gamma _{ \mathrm {c}}):=&\left \{{ (u,x) : \iota\left(u,x\right)\le \gamma _{ \mathrm {c}} }\right \},
\\ 
\mathcal {T}_d(D):=&\left \{{ (x, \hat{x}) : d(x, \hat{x})\le D }\right \},\\
\mathcal {T}_g(G):=&\left \{{ (\vec{u}, \vec{y}) : \Exp{\ell(\tilde{X}, \hat{f}^{(n)}(\vec{z},\Tilde{Y
}))}{\tilde X \tilde Y}\le G }\right \}, \label{eg:Tg}
\end{align}
where $\iota$ is the information density defined in ~\eqref{eq:def_information_density}, $\gamma_p$, $\gamma_c$ are predefined thresholds, and $G, D$ are the generalization error and distortion constraint, respectively. 
The first three sets already appeared in~\cite{watanabe2015nonasymptotic} for the conventional setup of data reconstruction with side information, while we introduce the last one specifically for the analysis of the generalization error of the regression problem. 

Accordingly, we define the information-density-distortion-generalization error vector as follows: 
\begin{align}
\label{eq_idl}
    \vec{i}(X, \vec{U}, \vec{Y}, \hat{X}):=\begin{bmatrix} -\iota\left(U , Y\right) \\ \iota(U,X)\\ d(X, \hat{X}) \\ \Exp{\ell(\Tilde{X}, \hat{f}^{(n)}(\vec{Z},\Tilde{Y})}{\tilde X \tilde Y} \end{bmatrix} ,
\end{align}
where $\vec{U}$ represents the full training sequence of length $n$, while $U$ refers to one variable within this sequence. The same applies for $\vec{Y}$ and $Y$.
Taking the expectation over the distribution $P_{X\vec{U}\vec{Y}\hat{X}}$ of this vector gives 
\begin{align}
    \vec {J}(\vec{i}):=\Exp{\vec{i}(X,\vec{U},\vec{Y}, \hat{X})}{} =\begin{bmatrix} -I(U;Y)\\ I(U;X) \\ \Exp{d(X, \hat{X})}{} \\ \Exp{\ell(\tilde{X}, \hat {f}^{(n)}(\vec{Z}, \tilde{Y})}{\Vec{Z}\tilde X \tilde Y} \end{bmatrix}\!,
\end{align}
where the sum of the first two components provides the Wyner-Ziv coding rate. The covariance matrix of this vector is defined as
\begin{align}
\label{eq_covariance_mtx}
    \mat{V} = \mathbb{C}\left( \vec{i}(X, \vec{U}, \vec{Y}, \hat{X})\right) .
\end{align}
Let $\mat{V}\in\mathbb {R}^{4\times 4}$ be a positive-semi-definite matrix. Given a Gaussian random vector $\vec{B}\sim \mathcal{N}(0,\mat{V})$, the dispersion region is defined with respect to the covariance matrix as\cite{tan_three_nit}
\begin{align}
\label{dispersion}
    \mathscr{S}( \mat{V}, \varepsilon ):= \{ \vec {b}\in \mathbb {R}^{4}: \Pr ( \vec {B}\le \vec {b})\ge 1- \varepsilon \}.
\end{align}

\subsection{Non-asymptotic regions}

The non-asymptotic achievability regions can be obtained by the method of channel resolvability \cite{hayashi2006general, cuff2013distributed}, or by mutual covering lemmas proposed in \cite{verdu}. In our case, we consider the former analysis and its extension to the case with side information~\cite{watanabe2015nonasymptotic}. In our proofs, we adapt the analysis of~\cite{watanabe2015nonasymptotic} by further considering the regression problem through the generalization error, and by taking into consideration the fourth set $\mathcal {T}_g(G)$ in~\eqref{eg:Tg}. This led to the following two Theorems.  
\begin{theorem}[Non-asymptotic achievable code]
\label{thm_achivability_excess_bound}
For arbitrary constants $\gamma_p,\gamma_c, D, G \geq0$, and positive integer N, there exists an $\left(n, M, D, G, \varepsilon \right)$ code satisfying
\begin{align} 
\varepsilon &\le \mathbb{P}_{X\vec{U}\vec{Y} \hat {X}} \big [(u,y) \in \mathcal {T} _{ \mathrm {p}}(\gamma _{ \mathrm {p}})^{c} \cup (u,x) \in \mathcal {T} _{ \mathrm {c}}(\gamma _{ \mathrm {c}})^{c}  
\cup (x,\hat{x}) \in \mathcal {T} _{ \mathrm {d}}( {D})^{c} \cup (\vec{u}, \vec{y})\in \mathcal {T}_g (G)^{c}   \big] \notag\\
&+ \frac {N}{2^{\gamma _{ \mathrm {p}}}| \mathcal {M}| } + \frac {1}{2}\sqrt {\frac {2^{\gamma _{ \mathrm {c}}}}{N}}. 
\end{align}
\end{theorem}
\begin{proof}
    The proof is provided in appendix \ref{apx:excess_prob}.
\end{proof}

By choosing $\gamma_p = \log \frac{N}{|\mathcal{M}_n|} + \log n$ and $\gamma_c = \log N - \log n $, and by applying Theorem \ref{thm_achivability_excess_bound} together with the multidimensional Berry-Esséen Theorem, we derive the achievable second-order rate-distortion-generalization error region as follows. 
\begin{theorem}[Second-order coding rate]
\label{thm_second_order_bound}
    For every $0<\varepsilon<1$, and $n$ sufficiently large, the $(n , \varepsilon)$-rate-distortion-generalization error function satisfies:
    \begin{align}
    \label{eq_non-asym-r}
        R(n, \varepsilon,D, G)\le&\inf \bigg \{\vec {M} \left ({ \vec{J}+ \frac { \mathscr {S}( \mat {V}, \varepsilon ) }{\sqrt {n}} + \frac {2\log n}{n} \vec {1}_{4}}\right )\bigg \},
    \end{align}
    with $\vec{M} = [1\quad 1\quad 0\quad 0]$.
\end{theorem}
\begin{proof}
The proof is provided in appendix \ref{apx:second_order_coding_rate}.
\end{proof}

The previous result is not a straightforward extension of the proofs in~\cite{watanabe2015nonasymptotic} as we introduce the generalization error term in the information-density-distortion-generalization error vector $\vec{i}$.  Especially, the vector $\vec{i}$ depends on the full sequence $\mathbf{Z}$ (it is not single-letter anymore), because the generalization error depends on the full training sequence. Therefore, the Berry-Esséen Theorem needs to be applied by conditioning on the other $n-1$ training samples.

Finally, the bound in Theorem~\ref{thm_second_order_bound} is composed by two parts. The vector $\vec{J}$ corresponds to the asymptotic result of Section~\ref{subsec:asymptotic_tradeoff}, while the other terms provide the non-asymptotic penalty introduced by the Gaussian approximation. 

\subsection{Non-asymptotic trade-off}
As in the asymptotic regime, we now investigate the trade-off between distortion and generalization error. To proceed, we further investigate the dispersion region to explore the relation between regression and reconstruction.

\begin{corollary}[Non-asymptotic trade-off for parametric and kernel regression]
\label{cor_non_aymp}
    For a finite sequence $(\vec{X}, \vec{Y})$ following the model in~\eqref{regression_model},  $0<\varepsilon<1$ and $n$ sufficiently large, there exists an achievable rate-distortion-generalization error region such that 
    \begin{align}
         R_{ b}(n, G, D, \varepsilon) > \max \{R_{ {b}}(n, G, \varepsilon), R_{b}(n, D, \varepsilon)\},
    \end{align}
    where $R_{ b}(\cdot)$ denotes the minimum achievable rate introduced by the right hand side of \eqref{eq_non-asym-r}.
\end{corollary}

\begin{proof}
    The proof is provided in Appendix \ref{apx:cor2}.
\end{proof}


The previous result shows that for the considered achievability scheme, there is no trade-off in terms of coding rate between the distortion and the generalization error.  
In order to prove Corollary~\ref{cor_non_aymp}, we first showed that the terms $d(X, \hat{X})$ and $\Exp{\ell(\Tilde{X}, \hat{f}^{(n)}(\vec{Z},\Tilde{Y})}{\tilde X \tilde Y}$ in~\eqref{eq_idl} are decorrelated, which turns into conditional independence with respect to the first two terms of the matrix, due to the Gaussian approximation from the Berry-Esséen Theorem.  
%
However, there is no guarantee that we can achieve the minimum for both criterion in finite block-length because of the excess probability constraint. 
Note also that this result is specific to the considered achievable coding scheme.  However, the approach of analyzing the correlation between terms in the information-density vector could be applied to other achievability schemes. 
Finally, the analysis in Appendix \ref{apx:cor2} highly depends on the Gaussian test channel we have chosen: the assumption that quantization error $\Phi$ and the system noise $N$ are independent from the source $X$ and $Y$ plays a vital role in the calculation of the correlation. 





\section{Numerical results} \label{sec:simuls}
\begin{figure*}
    \begin{subfigure}{.5\textwidth}
    \centering
    \includegraphics[width=\linewidth]{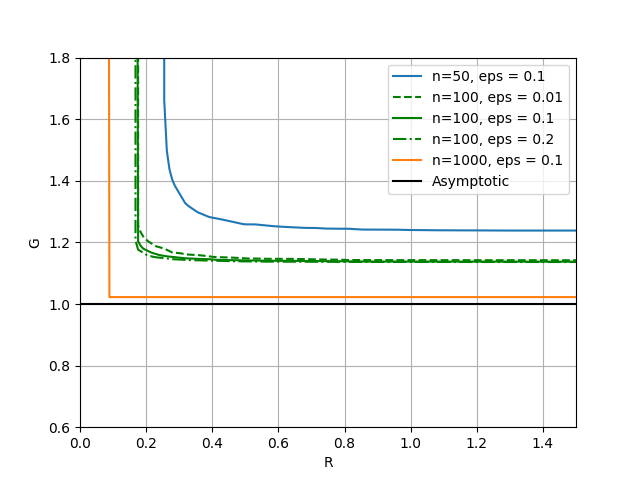}
    \caption{Rate-generalization error region for polynomial regression labeled on the block-length $n$ and the excess loss probability $\varepsilon$.}
    \label{fig:rg_poly}
    \end{subfigure}
    \begin{subfigure}{.5\textwidth}
    \centering
    \includegraphics[width=\linewidth]{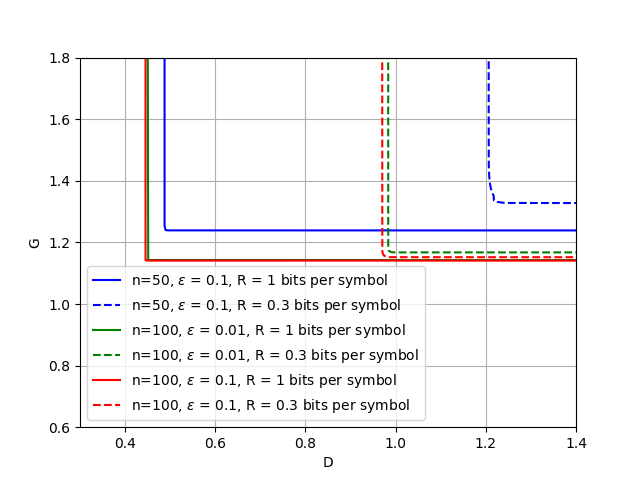}
    \caption{Distortion-generalization error region for polynomial regression on the block-length $n$, the excess loss probability $\varepsilon$ and rate $R$.}
    \label{fig:dg_poly}
    \end{subfigure}

    \begin{subfigure}{.5\textwidth}
    \centering
    \includegraphics[width=\linewidth]{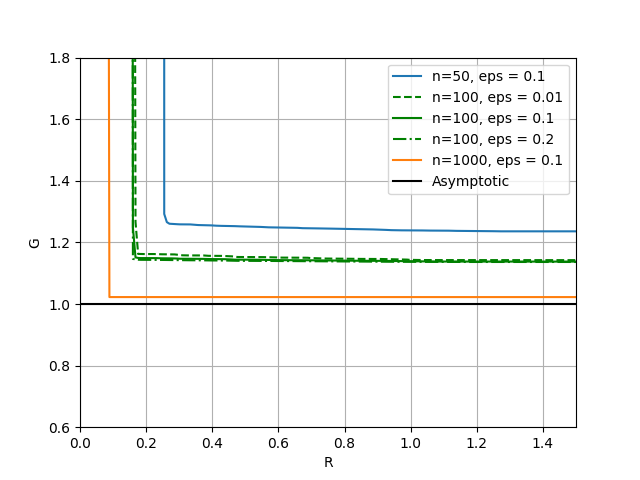}
    \caption{Rate-generalization error region for kernel regression labeled on the block-length $n$ and the excess loss probability $\varepsilon$.}
    \label{fig:rg_kernel}
    \end{subfigure}
    \begin{subfigure}{.5\textwidth}
    \centering
    \includegraphics[width=\linewidth]{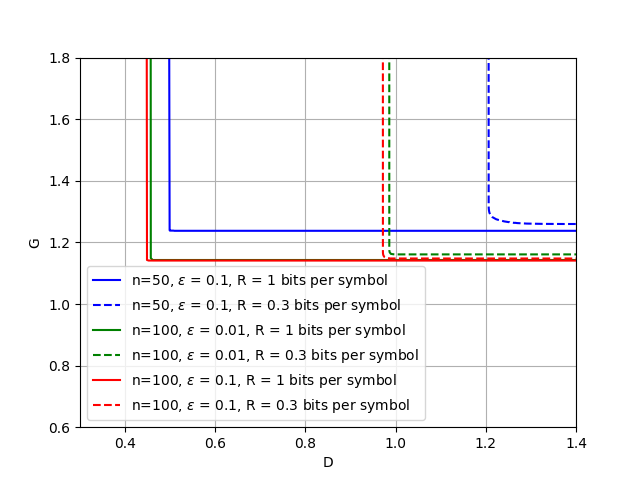}
    \caption{Distortion-generalization error region for kernel regression on the block-length $n$, the excess loss probability $\varepsilon$ and rate $R$. }
    \label{fig:dg_kernel}
    \end{subfigure}
    \caption{Non-asymptotic rate-distortion-generalization error region }
\end{figure*}

In this section, we provide numerical evaluations of the finite-length  achievable rate-distortion-generalization error regions provided in Section~\ref{sec:fbl}. As a particular case, we consider that $X$ and $Y$ follow a polynomial relation defined by $X = \vec{\beta}^T \vec{Y^*} + N$, where $\vec{\beta} = [2, 1, 1]^T $, $\vec{Y^*} = [Y^0, Y^1, Y^2]$, $\sigma = 1$, and $Y$ is uniformly distributed over $[-1, 1]$. We provide the finite-length achievable regions obtained from Theorem~\ref{thm_second_order_bound} for both parametric and non-parametric kernel regression.  In the latter case, we consider a Gaussian kernel, and the bandwidth $h_n$ for different block-length $n$ is set as $h_n = \left(\frac{(\sigma^2 + \sigma_\Phi^2)C_2}{C_1}n \right )^{-\frac{1}{5}}$. This value is known to be optimal in the asymptotic regime~\cite{wasserman_all_np}, but it might be sub-optimal for minimizing the expected generalization error at finite length. 

In both parametric and kernel regressions, the covariance matrix $\mat{V}$ defined in equation~\eqref{eq_covariance_mtx} has to be estimated. To do so, we sample information-density-distortion-generalization error vectors $\vec{i}$ in \eqref{eq_idl} by first generating $n$ samples of $X$ and $Y$, and then calculating the four components of the vector. 
In order to do so, we need the following prior results: 

    \paragraph*{ The probability density function of $U$}by the Theorem of variable change, for $\beta_2 > 0$ and $\beta_1^2 + 4\beta_2(w-\beta_0)\geq0$, we can show that the distribution of $W=\vec{\beta}^T \vec{Y}^\star$ is given by: 
    \begin{equation}
        P_W(w) = \begin{cases}
            \frac{1}{\sqrt{\beta_1^2 + 4\beta_2(w-\beta_0)}} & |y_1(w)| \le 1 \ \text{and}\  |y_2(w)| \le 1,\\
            \frac{1}{2\sqrt{\beta_1^2 + 4\beta_2(w-\beta_0)}}  & |y_1(w)| \le 1 \ \text{or}\  |y_2(w)| \le 1,\\
            0 &  \text{otherwise},
        \end{cases}\notag
    \end{equation}
    where $y_1 = \frac{-\beta_1 - \sqrt{\beta_1^2 + 4\beta_2(w-\beta_0)}}{2\beta_2}, y_2 = \frac{-\beta_1 + \sqrt{\beta_1^2 + 4\beta_2(w-\beta_0)}}{2\beta_2}$. The probability density function of $U = \alpha (W + N + \Phi)$ can then be expressed as 
    \begin{align}
        P_U(u) = \frac{1}{\alpha\sqrt{2\pi(\sigma^2 + \sigma_\Phi^2)}}\int_{-\infty}^{\infty} P_W(w) e^{-\frac{(\frac{u}{\alpha}-w)^2}{2(\sigma^2+\sigma_\Phi^2)}}dw,
    \end{align}
    which can be evaluated numerically;
    \paragraph*{ The conditional distribution of $(U|X)$ and $(U|Y)$}by the test channel defined in Section~\ref{subsec_proof}, we have $(U|Y) \sim \mathcal{N}(0, \alpha^2(\sigma^2 + \sigma_\Phi^2))$ and $(U|X) \sim \mathcal{N}(0, \alpha^2\sigma_\Phi^2)$.

Then, the main steps of the numerical evaluation of the covariance matrix $\mat{V}$ are as follows:
\begin{enumerate}
    \item Generate $n$ samples 
    of $X, Y$ and $U$, according to the Gaussian test channel defined in Section \ref{subsec_proof}
    \item For each sample $(u,x,y)$, the information densities $\iota(x;u)$ and $\iota(u;y)$ are obtained with \eqref{eq:def_information_density}; 
    \item The distortion is calculated by: 
    \begin{align}
        d(x, \hat{x}) = \left(\hat{x} - x \right)^2.
    \end{align}
    \item The generalization error $G(\hat{f}^{(n)}, \vec{Z})$ is given by \eqref{eq_generalization_error}, where the expectation is estimated with $N^\star = 500$ samples for kernel regression, and is directly calculated by \eqref{eq_generalization_parametric} for parametric regression. 
    \item Repeat steps $1)$ to $4)$ $N^\star=500$ times to get numerical estimation of the covariance matrix \eqref{eq_covariance_mtx}. 
\end{enumerate}
Then the achievable region is obtained by Theorem~\ref{thm_second_order_bound}. 

In addition, Figures~\ref{fig:rg_poly} and ~\ref{fig:rg_kernel} show the boundaries of the rate-generalization error regions for polynomial and kernel regressions, considering different block-length $n$ and excess probability $\varepsilon$. In both cases, we observe that the achievable regions converges to the asymptotic one as $n$ increases, and we also observe that lower rates can be achieved if higher excess probabilities are allowed.
Figures \ref{fig:dg_poly} and \ref{fig:dg_kernel} illustrate the distortion-generalization error region for coding rates $R=0.3$ bit/symbol and $R=1$ bit/symbol. The regions are consistent with our Corollary~\ref{cor_non_aymp} which states that the decorelation between the distortion and the generalization error results in the absence of trade-off between the two criterions. In addition, we observe that both distortion and generalization error decrease with the coding rate $R$.

Finally, for fixed rate $R$ and excess probability $\epsilon$, we see that the generalization error of OLS estimator converges faster than the generalization error of kernel estimators, which is consistent with the different convergence rates of these two types of regression. Especially it is shown in~\cite{wasserman_all_np} that for kernel regression, the optimal $h$ is of order $O\left(n^{-\frac{1}{5}}\right)$ and that the expected generalization error decreases to the minimum expected loss $\sigma^2$ at rate $O\left(n^{-\frac{4}{5}}\right)$. On the opposite, in  parametric methods, the generalization error decreases to $\sigma^2$ at rate $O\left(n^{-1}\right)$. The slower rate $O\left(n^{-\frac{4}{5}}\right)$ is the price of using non-parametric methods.

\section{Conclusion} \label{sec:conclusion}
In this article, we investigated regression under the generalization error criterion within the framework of goal-oriented communications. Our information-theoretic analysis provided rate-generalization error regions for parametric regression and kernel regression in both asymptotic and non-asymptotic regimes. We improved upon existing bounds~\cite{raginsky} in the asymptotic regime, demonstrating convergence of generalization error to the minimum expected loss. 
In the non-asymptotic regime, we relied on the finite-length tools introduced in~\cite{watanabe2015nonasymptotic} and extended these tools to our regression problems. We further investigated the trade-off between regression and reconstruction, and as a key finding of our research, we showed that, in both cases (asymptotic and non-asymptotic), there is no trade-off between reconstruction and regression. A posterior remark of this result is that for both reconstruction and regression, we used the same test-channel. 
The established optimality of this test channel in infinite block-length further solidified our findings. The converse in the non-asymptotic regime remains an open question, inviting further exploration in future works.

\appendices

\section{Preliminary Theorems}
\label{apx:prelimenary}

Here we restate the channel resovability problem~\cite[Chapitre 6]{han_information_spectrum} and related definitions used in~\cite{watanabe2015nonasymptotic}. The statements of~\cite{watanabe2015nonasymptotic} apply for discrete source, and this appendix generalizes them to arbitrary distributions. 


\subsection{Smoothing of a distribution}

Denote $\mathscr{P}(\set{X})$ as the set of all probability distributions on a measurable space $\left(\set{X} , \mathscr{B}\left(\set{X}\right)\right)$, and let $\mathscr{P}'(\mathcal{X})$ be the set of all sub-normalized non-negative functions (not necessarily a probability measure). Note that if $P\in \mathscr{P}'(\mathcal{X})$ is normalized then $P \in \mathscr{P}(\mathcal{X})$. For a subset $\mathcal{T} \subset\mathcal{X}$, the smoothed sub-normalized function $\Bar{P}_X$ of $P_X$ is defined as, $\forall A\in \mathscr{B}\left(\set{X}\right)$
\begin{align}
    \Bar{P}_X(A) =  \int_{A} \mathbf{1}[x \in \mathcal{T}] p_X(x)dx.
\end{align}
For two functions $P, Q \in \mathscr{P}'(\mathcal{X})$, the variational distance between $P$ and $Q$ is:
\begin{equation}
    d_{TV}(P, Q) = \sup_{A \in \mathscr{B}\left(\set{X}\right)} \left| P(A) - Q(A)\right|,
\end{equation}
and it has the following property. 
    \begin{lemma}[Property of variational distance\cite{watanabe2015nonasymptotic}]
        For a distribution $P\in \mathscr{P}(\mathcal{X})$ and a sub-normalized measure $Q \in \mathscr{P}'(\mathcal{X})$, and any subset $\Gamma$ of $\mathcal{X}$,
        \begin{equation} 
        \label{eq_lemma_tv}
        P(\Gamma )\leq Q(\Gamma )+d_{TV}(P,Q)+\frac {1-Q( \mathcal {X})}{2}. 
        \end{equation}
    \end{lemma}
Hence the variational distance between the original distribution and a smoothed one is
\begin{align}
    d_{TV}(P, \Bar{P}) = \frac{P(\mathcal{T}^c)}{2},
\end{align}
where $\set{T}^c$ stands for the complementary set of $\set{T}$. For a channel $P_{U|X}:\mathcal{X}\rightarrow\mathcal{U}$ a subset $\mathcal{T} \subset \mathcal{X} \times \mathcal{U}$, and the event $B \in \mathscr{B}\left(\set{U}\right) $ and $x\in \mathcal{X}$, the smoothed conditional function $\Bar{P}_{U|X}$ is defined by
\begin{align}
    \Bar{P}_{U|X}(B|X = x) = \int_{B} \mathbf{1}[(u, x) \in \mathcal{T}]p_{U|X}(u|x)du .
\end{align}

\subsection{Channel resolvability and identification code}

Let us consider a channel $P_{U|X}$ and an input distribution $P_X$.
In the channel resolvability problem, we choose $M$ elements in the input set $\mathcal{X}$, \emph{i.e.}, a codebook $\mathcal{C} = \{ x_1, x_2, ..., x_M\}$, such that the output distribution $P_U$ expressed from the input distribution $P_X$ as
\begin{align}
    P_U(B) = \int_B\int_\set{X} p_X(x) p_{U|X}(u|x) dudx
\end{align}
is close enough to the output distribution $P_{U'}(B)$ obtained when the input is assumed to be uniformly distributed \cite{hayashi2006general, cuff2013distributed}, i.e.
\begin{align}
    P_{U'}(B) = \int_B\sum_{i=1}^M \frac{\mathbf{1}[x=x_i]}{M} p_{U|X}(u|x)du,
\end{align}
where we suppose the channel $P_{U|X}$ is absolutely continuous. 
By the soft covering lemma from \cite[Corollary 7.2]{cuff2013distributed} and \cite[Lemma 2]{hayashi2006general}, the following result states that
\begin{corollary}[Lemma 25 of \cite{watanabe2015nonasymptotic}]
    Let $\mathcal{T} = \mathcal {T} _ \mathrm {c}(\gamma _ \mathrm {c})$ defined in \eqref{eq_tc}, for any $\gamma_c \geq 0$, we have 
    \begin{align}
        \Exp{d_{TV}(\bar{P}_U, \bar{P}_{U'})}{\mathcal{C}} \leq \frac {\Delta (\gamma _{ \mathrm {c}}, P_{UX})}{2\sqrt {M}},
    \end{align}
    with $\Delta (\gamma _{ \mathrm {c}}, P_{UX}) = \sqrt{\Exp{\frac{dP_{U|X}(u|x)}{dP_U(u)}\mathbf {1}[ (u,x) \in \mathcal {T} _ \mathrm {c}(\gamma _ \mathrm {c})]}{UX}}$.
\end{corollary}

\section{Gaussian test channel and  coding scheme}
\label{apx:test_channel}
Consider the test channel $ U = \alpha(X + \Phi)$ defined in Section~\ref{subsec_proof}. 
Since we assume that the function $f$ and  the joint distribution $P_{XY}$ are unknown in both the encoder and the decoder, we employ the achievable scheme proposed in \cite{draper2004universal} based on the method of types and binning. However, compared to~\cite{draper2004universal}, no incremental transmission is needed since we suppose that the noise variance $\sigma^2$ is known. Therefore, the test channel parameters as well as the binning rate are fixed. In fact, by setting
\begin{align}
    \alpha = \frac{\sigma^2 - D}{\sigma^2} \text{ and } \sigma^2_{\Phi} = \frac{D\sigma^2}{\sigma^2 - D}
\end{align}
the distortion constraint $\Exp{d(X, \hat{X})}{} \le D$ can be achieved for Gaussian source~\cite{wyner1978}.
Thus the variable-rate scheme in \cite{draper2004universal} becomes a fixed rate coding scheme. However, we need to keep the prefix transmission of types applied by Draper\cite{draper2004universal} since the joint distribution $P_{XY}$ is unknown.

This scheme works as follows:
\begin{enumerate}
    \item The codebook is formed by generating randomly $2^{nR_1}$ sequences $\vec{u}$, which are uniformly distributed into $2^{nR}$ bins, with $R_1 > R$.
    \item The encoder identifies a sequence $\vec{u}$ which is typical with $\vec{x}$, and transmits the index of the bin to which $\vec{u}$ belongs.
    \item At the decoder, a typicality test is performed between the side information $\vec{y}$ and all the sequences in the bins, allowing a sequence $\vec{\hat{u}}$ to be extracted from the bin. 
\end{enumerate}
Draper shows in \cite{draper2004universal} that the probability of debinning error can be made as small as desired if the block length $n$ is large enough.
In addition, given that $D<\sigma_x^2$ and $(X\left|\right. Y)$ is Gaussian, we will show that the rate-distortion function $R_b(D) = \frac{1}{2} \log \left(1 + \frac{\sigma^2}{\sigma_\Phi^2}\right)$ is achievable for $\Exp{d(X, U)}{XU} \leq D$, where $D$ is a function of $\sigma_{\Phi}^2$.
Next, in our proof, we need to express the generalization error for regression, and to analyze its convergence with respect to $n$. In our proofs, this analysis is specific to the considered regression problem, parametric or non-parametric. For parametric regression, this analysis is provided in Section~\ref{subsec_proof}. For kernel regression, the analysis is provided in the next Appendix.  

\section{Proof of Theorem~\ref{thm_kernel}}
\label{apx:kernel}
Consider the definition of $\hat{f}(y)$ in equation \eqref{eq_kernel_uy}. For a given $\Tilde{y}$, for $\forall i \in \llbracket 1,n\rrbracket$, note that
\begin{align}
    \frac{U_i}{\alpha} &=  f(Y_i) + N_i + \Phi_i
    = f(\tilde y) + (f(Y_i) - f(\tilde y)) + (N_i + \Phi_i) .
\end{align}
Therefore,
\begin{align}
\label{eq_apxc_1}
    \frac{1}{nh}\sum_{i=1}^n K\left(\frac{\tilde y-Y_i}{h}\right)\frac{U_i}{\alpha} &= \frac{1}{nh}\sum_{i=1}^n K\left(\frac{\tilde y-Y_i}{h}\right)f(\tilde y) 
    +\frac{1}{nh}\sum_{i=1}^n K\left(\frac{\tilde y-Y_i}{h}\right)(f(Y_i) - f(\tilde y)) 
    \notag\\
    & \quad +\frac{1}{nh}\sum_{i=1}^n K\left(\frac{\tilde y-Y_i}{h}\right)(N_i + \Phi_i) \notag\\
    &= \hat{p}_Y(\tilde y)f(\tilde y) + \hat{m}_1(\tilde y) + \hat{m}_2(\tilde y)
\end{align}
where $\hat{p}_Y(\tilde y) = \frac{1}{nh}\sum_{i=1}^n K\left(\frac{\tilde y-Y_i}{h}\right)$ is the kernel density estimation of $\tilde{y}$ from observation $\vec Y$\cite{wasserman_all_np}, $\hat{m}_1(\tilde y) = \frac{1}{nh}\sum_{i=1}^n K\left(\frac{\tilde y-Y_i}{h}\right)(f(Y_i) - f(\tilde y))$ and $\hat{m}_2(\tilde y) = \frac{1}{nh}\sum_{i=1}^n K\left(\frac{\tilde y-Y_i}{h}\right)(N_i + \Phi_i)$. The analysis of the asymptotic distribution of the kernel estimator $\hat{f}(\Tilde{y})$ is based on \cite{hardle2004nonparametric}, which relies on the analysis of the numerator and the denominator of \eqref{eq_kernel_uy}.

First, for $\hat{m}_2(\tilde y)$, we have that $\Exp{\hat{m}_2(\tilde y)}{Y\Phi N } = 0$, and its variance can be expressed as: 
\begin{align}
    \Var{\hat{m}_2(\tilde y)}{} &= \Var{\frac{1}{nh}\sum_{i=1}^n K\left(\frac{\tilde y-Y_i}{h}\right)(N_i + \Phi_i)}{} \\ 
    &=\frac{\sigma^2 + \sigma_\Phi^2}{nh^2} \int K^2\left(\frac{\tilde y-y}{h}\right) p_Y(y)dy \\
    &=\frac{\sigma^2 + \sigma_\Phi^2}{nh}\int K^2(u) p_Y(\tilde y + uh) du \label{pr_change_of_measure}\\
    &= \frac{\sigma^2 + \sigma_\Phi^2}{nh}\int K^2(u) \left( p_Y(\tilde y) + p_Y'(\tilde y)uh + o(h)\right) du \label{pr_taylor}\\
    &=\frac{(\sigma^2 + \sigma_\Phi^2)p(\tilde y)}{nh}\int K^2(u)du + o\left(\frac{1}{nh}\right)
\end{align}
where \eqref{pr_change_of_measure} follows from a change of variable, and~\eqref{pr_taylor} comes from  a Taylor approximation of $p_Y(\tilde y+uh)$ when $h\rightarrow 0$.

Also, following the same derivation as in \cite{hardle2004nonparametric}, for $\hat{m}_1(\tilde y)$, we show that
\begin{align}
\label{eq_apxc_2}
    \Exp{\hat{m}_1(\tilde y)}{} 
    &=\frac{h^2}{2}\left(2f'(\tilde y)p_Y'(\tilde y) + f''(\tilde y)p_Y(\tilde y)\right)\int u^2K(u)du 
    + o(h^2)
\end{align}
and $\Var{\hat{m}_1(\tilde y)}{} = O(\frac{h}{n})$ 
, which is negligible compared to the variance of $\hat{m}_2(\tilde y)$. 
From \eqref{eq_apxc_1} to \eqref{eq_apxc_2}, the central limit theorem is applied to obtain that as $h\rightarrow0$ and $nh\rightarrow\infty$, we have
\begin{align}
\label{eq_apxc_3}
    &\hat{m}_1(\tilde y) \xrightarrow{p} \frac{h^2}{2}\left(2f'(\tilde y)p_Y'(\tilde y) + f''(\tilde y)p_Y(\tilde y)\right)\int u^2K(u)du\\
    &\hat{m}_2(\tilde y) \xrightarrow{d} \mathcal{N}\left(0, \frac{(\sigma^2 + \sigma_\Phi^2)p_Y(\tilde y)}{nh}\int K^2(u)du\right)
\end{align}
where $\xrightarrow{p}$ denotes the convergence in probability and $\xrightarrow{d}$ denotes the convergence in distribution. By the property of kernel density estimation \cite{wasserman_all_np}, it can be shown that $\hat{p}_Y\xrightarrow{p} p_Y$.
Next, the kernel function \eqref{eq_kernel_regression} can be expressed as : 
\begin{align}
\label{eq_apxc_4}
    \hat{f}(\tilde y) = f(\tilde y) + \frac{\hat{m}_1(\tilde y)}{\hat{p}_Y(\tilde y)} + \frac{\hat{m}_2(\tilde y)}{\hat{p}_Y(\tilde y)}
\end{align}

By equation \eqref{eq_apxc_3} to \eqref{eq_apxc_4}, we have the bias and variance in \eqref{eq_bn} and \eqref{eq_vn}. By Slutsky's theorem~\cite{goldberger}, we have : 
\begin{align}
    \frac{\hat{m}_1(\tilde y) + \hat{m}_2(\tilde y)}{\hat{p}_Y(\tilde y)} \xrightarrow{d} \mathcal{N}\left(\frac{\Exp{\hat{m}_1(\Tilde{y}}{}}{p_Y(\tilde y)}, \frac{\Var{\hat{m}_2(\Tilde{y}}{}}{p_Y(\tilde y)}\right).
\end{align}
Hence
\begin{align}
    &\hat{f}(\tilde y) - f(\tilde y) \xrightarrow{d} \mathcal{N}\left(b_n(\Tilde{y}), V_n(\Tilde{y})^2\right).
\end{align}
where $b_n(\Tilde{y}), V_n(\Tilde{y})$ are defined in \eqref{eq_bn}, \eqref{eq_vn}. According to \eqref{eq_exp_ge}, the generalisation error can be expressed as:  
\begin{align}
    \Exp{G(\hat{f}^{(n)}, \vec{Z})}{\vec{Z}} &= \sigma^2 + \frac{h^4}{4}C_1 + \frac{\sigma^2 + \sigma_\Phi^2}{nh} C_2 
    + o\left(\frac{1}{nh}\right) + o(h^4)
\end{align}
where $C_1 = \int \left(2\frac{f'(y)p_Y'(y)}{p_Y(y)} + f''(y)\right)^2 dy \left(\int u^2K(u)du \right )^2$ and $C_2 = \int \frac{1}{p_Y(y)} dy\int K^2(u)du$. 
Recall that the asymptotic generalization error with uncompressed observations $( \vec X,  \vec Y)$ is \cite{wasserman_all_np}
\begin{align}
    \Exp{G(\hat{f}^{(n)}, \vec{XY})}{\vec{XY}} &=\sigma^2 + \frac{h^4}{4}C_1 + \frac{\sigma^2 }{nh} C_2 
     + o\left(\frac{1}{nh}\right) + o(h^4)
\end{align}
which indicates that asymptotically, the regression performed on coded data can achieve the same performance than that the one performed on original data. 

\section{Proof of Corollary \ref{corollary_asymptotic}}
\label{apx:corollary_1}
We first consider the conditional setup where the side information $Y$ is available to both the encoder and the decoder. In this case, for $(X|Y) \sim \mathcal{N}(0, \sigma^2)$, the optimal conditional rate-distortion function is~\cite{gray1972conditional}
\begin{align}
\label{eq:rd_cond}
R_{X|Y}(D) = \frac{1}{2}\log \left( \frac{\sigma^2}{D} \right) ,
\end{align}
We now show that when the same relation between $X$ and $Y$ holds, i.e. $(X|Y) \sim \mathcal{N}(0, \sigma^2)$, but the side information is only available at the decoder, the Wyner-Ziv rate-distortion function $R_{WZ}(D)$ is equal to the conditional one $R_{X|Y}(D)$. Using the test channel provided in previous section $U = \alpha(X+\Phi)$ with $\alpha = \frac{\sigma^2 - D}{\sigma^2}$ and $\sigma^2_{\Phi} = \frac{D\sigma^2}{\sigma^2 - D}$, together with the scheme proposed in \cite{draper2004universal}, the intermediate variable $U$ can be recovered without error as the block-length tends to infinity. By considering the minimum mean square error estimator for the reconstruction $\hat{X}$ of $X$, it can be show that $\Exp{(X-\hat{X})^2}{}  = (\alpha-1)^2\sigma^2 + \alpha^2 \sigma_{\Phi}^2$. Replacing $\alpha$ and $\sigma_\Phi$ by their expression leads to $\Exp{(X-\hat{X})^2}{} = D$. 

Second, the binning rate in \cite{draper2004universal} can be expressed as 
\begin{align}\notag
    I(X;U) - I(Y;U) &=  h(U|Y) - h(U|X) \\
    &= \frac{1}{2} \log \left(\frac{\sigma^2 + \sigma_{\Phi}^2}{\sigma_{\Phi}^2}\right)
\end{align}
where $h(\cdot)$ denotes the differential entropy and the last equality comes from the fact that under the previous test channel both $(U|Y)$ and $(U|X)$ follow a Gaussian distribution. 
By replacing $\sigma_\Phi^2$ in the previous expression, we obtain $R_{WZ}(D) = R_{X|Y}(D)$, the optimal rate for reconstruction. Then by Theorem~\ref{thm_parametric}, for any positive $R_{WZ}(D)$, the pair $\left(R_{WZ}(D),0\right)$ is also achievable for the generalization error using the same scheme. In other words, it states that $R(D, G) = R_{WZ}(D)$ for all $G\geq \sigma^2$.
This indicates that the Gaussian test channel is also optimal in our regression setup as long as the conditional distribution of $(X|Y)$ is Gaussian, whatever the distribution of $Y$.


\section{Proof of theorem \ref{thm_achivability_excess_bound}}
\label{apx:excess_prob}
This proof follows the channel resolvability type code of \cite{watanabe2015nonasymptotic} for the Wyner-Ziv problem. We adapt it so as to also account for the generalization error in the information-density vector defined in \eqref{eq_idl}. Some preliminary results for channel resolvability and identification code were provided in Appendix \ref{apx:prelimenary}.


\textit{Code construction}: The encoder uses a stochastic map $P_{I\left| X \right.}: \set{X} \rightarrow \set{I}$. It generates $i\in\mathcal{I}$ according to $P_{I\left|X\right. }$.
Then the encoder sends $i$ by random binning $\kappa : \mathcal{I}\rightarrow\mathcal{M} $. It means for every $i\in\mathcal{I}$, it is independently and uniformly assigned to a random bin $m\in\mathcal{M}$. For given $m\in\mathcal{M}$ and $y\in\mathcal{Y}$, the decoder finds the unique index $i\in\mathcal{I}$ such that $\kappa(i) = m$, and  
\begin{equation} 
\label{tp}
(u_{i},y)\in \mathcal {T} _ \mathrm {p}(\gamma _ \mathrm {p}). 
\end{equation}
As mentioned in the channel resolvability problem in Appendix \ref{apx:prelimenary}, the stochastic map $P_{I\left|X\right. }$ is constructed such that the joint distribution $P_{\hat{I} X}$ is indistinguishable from $P_{IX'}$, where for any measurable $E \in \set{P}(\mathcal{I}) \times \mathscr{B}(\mathcal{X})$
\begin{align}
    P_{IX'}(E) = \int_E \sum_i^{|\mathcal{I}|} \frac{\mathbf{1}[u_i = u]}{|\mathcal{I}|}p_{X|U}(x|u) dx ,
\end{align}
and $P_{I X'}$ is the joint distribution of the couple $\left(I,X'\right)$, where $X'$ is the random variable that induces the probability measure $P_{X\left|U\right.}$ when $U$ is chosen uniformly in the codebook $\left\{ u_1 , \cdots u_{\left|\set{I}\right|}\right\}$.
Then the decoder has two objectives: i) performing the regression according to $\vec{U}$ and $\vec{Y}$ to obtain a function $\hat{f}$, and ii) reconstruction of $X$.
In both cases, a decoder error occurs if there is no $i$ satisfying \eqref{tp} or if there is more than one such $i$ satisfying \eqref{tp}.

Let $\hat{I}$ be random index chosen by the encoder via the stochastic map $P_{I|X}$. The joint distribution of $(\hat{I}, X)$ is given by
\begin{equation} 
P_{ \hat {I} X} = P_{X} P_{I\left|X\right. }. 
\end{equation}
And the joint distribution of $(\hat{I}, X, Y, \hat{X}, G)$ is given by
\begin{align} 
P_{ \hat {I} X Y \hat {X}} = P_{ \hat {I} X} P_{Y|X} P_{ \hat {X}| U Y}.
\end{align}
Note that here the generalization error is fully determined by the sequence $\vec{U}$ and $\vec{Y}$.
The smoothed versions $\Bar{P}_{ \hat {I} X}$ and $\Bar{P}_{ \hat {I} X Y \hat {X}}$ are given by
\begin{align}
    &\Bar{P}_{\hat{I}X}(E) = \int_E \sum_i^{|\mathcal{I}|} \mathbf{1}[(u_i, x)\in \mathcal {T}_{ \mathrm {c}} (\gamma _{ \mathrm {c}})] p_X(x)p_{I|X}(i|x) dx \label{eq:pix}\\
    &\Bar{P}_{ \hat {I} X Y \hat {X}} = \Bar{P}_{ \hat {I} X} P_{Y|X} P_{ \hat {X}| U Y}.
\end{align}
$P_{e,n}(D, G)=\prob{\left\{d(X,\hat X) \geq D\right\} \ \cup \ \left\{G(\hat{f}^{(n)}, \vec{Z}) \geq G\right\}}{}$ from Definition \ref{def:nmdleps}. This error probability is bounded away from $0$ if at least of one of the following error events occurs:
\begin{align}
\mathcal {E}_{0}:=&\left \{ (\vec{u}, \vec{y})\notin \mathcal {T}_{ {g}} (G) \right \}, \\ 
\mathcal {E}_{1}:=&\left \{{ (x, \hat x)\notin \mathcal {T}_{ {d}}(D) }\right \}, \\ 
\mathcal {E}_{2}:=&\left \{{ (u_{i},y)\notin \mathcal {T} _ \mathrm {p}(\gamma _ \mathrm {p})}\right \}, \\ 
\mathcal {E}_{3}:=&\left \{{ \exists \, i'\neq i ~\text {s.t. } \kappa ( i')= \kappa (i),(u_{ i'},y)\in \mathcal {T} _ \mathrm {p}(\gamma _ \mathrm {p})}\right \}\!. 
\end{align}
For fixed codebook $\mathcal{C}$, following the same step as in \cite{watanabe2015nonasymptotic}, the excess probability is thus upper bounded by : 
\begin{align}
&P_{ \hat {I} X Y \hat {X}} ( \mathcal {E}_{0}\cup \mathcal {E} _{1} \cup \mathcal {E} _{2} \cup \mathcal {E} _{3}) \notag \\
\leq&\bar {P}_{I X' Y \hat {X}} ( \mathcal {E}_{0}\cup \mathcal {E} _{1} \cup \mathcal {E} _{2}) + \bar {P}_{I X' Y \hat {X}}(\mathcal {E} _{3}) + d_{TV}(P_{ \hat {I} X }, \bar {P}_{I X'})  \notag \\
& + \frac {1- \bar {P}_{I X' Y \hat {X}}( \mathcal {I}\times \mathcal {X}\times \mathcal {Y}\times \hat { \mathcal {X}})}{2}
\end{align}

Next, the excess probability $P_{e,n}(D,G)$ is averaged over the random coding function $\kappa$ and the random codebook $\mathcal{C}$ can be upper bounded as 
\begin{align} 
\label{prob}
&\mathbb {E}_{ \kappa } \mathbb {E}_ \mathcal {C}[ P_{e,n}(D,G)]\notag\\
\leq&\mathbb {E}_{ \kappa } \mathbb {E}_ \mathcal {C}\left [{ P_{ \hat {I} X Y \hat {X}} ( \mathcal {E}_{0}\cup \mathcal {E} _{1}\cup \mathcal {E} _{2} \cup \mathcal {E} _{3}) }\right ]\quad \notag\\
\leq& \Exp{\bar {P}_{I X' Y \hat {X}} ( \mathcal {E}_{0}\cup \mathcal {E} _{1} \cup \mathcal {E} _{2})}{\mathcal{C}} + \mathbb {E}_{ \kappa }\Exp{\bar {P}_{I X' Y \hat {X}}(\mathcal {E} _{3})}{ \mathcal{C}}
\notag \\&
 + \Exp{d_{TV}(P_{ \hat {I} X }, \bar {P}_{I X'})}{\mathcal{C}}
+ \Exp{\frac {1- \bar {P}_{I X' Y \hat {X}}( \mathcal {I}\times \mathcal {X}\times \mathcal {Y}\times \hat { \mathcal {X}})}{2}}{\mathcal{C}} .
\end{align}
The expectation of the first term can be expressed as: 
\begin{align} 
 &\Exp{\prob{(u_{i},x)\in \mathcal {T} _ \mathrm {c}(\gamma _ \mathrm {c})\cap \big(\mathcal {E}_{0}\cup \mathcal {E} _{1} \cup \mathcal {E} _{2}\big)}{I X' Y\hat{X}}  }{\mathcal {C}} \\ \notag
&= \prob{(u,x)\in \mathcal {T} _ \mathrm {c}(\gamma _ \mathrm {c})\cap \big(\mathcal {E}_{0}\cup \mathcal {E} _{1} \cup \mathcal {E} _{2} \big )}{X\vec{U}\vec{Y} \hat{X}} .
\end{align}
For the three last terms as proved in \cite{watanabe2015nonasymptotic}, we can prove that the second term in \eqref{prob} is upper bounded as: 
\begin{align}
& \mathbb {E}_{ \kappa } \mathbb {E}_ \mathcal {C}\left [{\bar P_{I X' Y } ( \mathcal {E}_{3})}\right ] \notag \\
= & \mathbb {E}_{ \kappa } \mathbb {E}_ \mathcal {C}\bigg [\int_{\mathcal{UXY}}\sum _{i} \frac {\mathbf{1}\left[(u_i, x)\in \mathcal{T}_c(\gamma_c)\right]}{| \mathcal {I}|} 
\times \mathbf {1}[ \exists i'\neq i\  s.t. \kappa(i') = \kappa(i), (u_{ i'},y) \in \mathcal {T} _ \mathrm {p}(\gamma _ \mathrm {p})]\notag\\
& \quad p_{X|U}(x|u)p_{Y|X}(y|x)dx dy \bigg ] \\
\le& \mathbb {E}_{ \kappa } \mathbb {E}_ \mathcal {C}\bigg [\int_{\mathcal{UXY}}\sum _{i} \frac {\mathbf{1}\left[(u_i, x)\in \mathcal{T}_c(\gamma_c)\right]}{| \mathcal {I}|} 
| \mathcal {I}| \sum_{i'\neq i}\frac{\mathbf{1}[\kappa(i') = \kappa(i)] \times \mathbf {1}[ (u_{i'},y) \in \mathcal {T} _ \mathrm {p}(\gamma _ \mathrm {p})]}{| \mathcal {I}|}  \notag\\
&\quad p_{X|U}(x|u)p_{Y|X}(y|x)dx dy \bigg ] \\
\le&\frac {| \mathcal {I}|}{| \mathcal {M}|}\int_{\mathcal{UXY}} \mathbf{1}\left[(u, x)\in \mathcal{T}_c(\gamma_c)\right]  p_{XYU}(x, y, u)dxdydu 
\int_{\mathcal{U}} \mathbf {1}[ (u,y) \in \mathcal {T} _ \mathrm {p}(\gamma _ \mathrm {p})]p_U(u)du   \label{exp_k}\\
\le&\frac {| \mathcal {I}|}{| \mathcal {M}|}  \int_{\mathcal{U}\mathcal{Y}} \mathbf {1}[ (u,y) \in \mathcal {T} _ \mathrm {p}(\gamma _ \mathrm {p})] p_{ U }(u)p_Y(y)dudy   \label{exp_c}\\
\le&\frac {|\mathcal {I}|}{2^{\gamma_p}| \mathcal {M}|},
\end{align}
where \eqref{exp_k} comes from the fact that $\Exp{\mathbf{1}[\kappa(i') = \kappa(i)]}{\kappa}  \le \frac{1}{|\mathcal{M}|}$.
The third term can be upper bounded as 
\begin{align} 
&\mathbb {E}_ \mathcal {C}[ d_{TV}(P_{ \hat {I} X }, \bar {P}_{I X'})]\notag\\
\le&\mathbb {E}_ \mathcal {C}[ d_{TV}(P_{ \hat {I} X }, \bar {P}_{\hat {I} X})] + \mathbb {E}_ \mathcal {C}[ d_{TV}(\bar P_{ \hat {I} X }, \bar {P}_{I X'})]\\
\le&\frac { P_{UX} ( (u,x) \notin \mathcal {T} _ \mathrm {c}(\gamma _ \mathrm {c})) }{2}\notag + \frac {\Delta (\gamma _{ \mathrm {c}}, P_{UX})}{2\sqrt {| \mathcal {I}|}}\\
\le&\frac { P_{UX} ( (u,x) \notin \mathcal {T} _ \mathrm {c}(\gamma _ \mathrm {c})) }{2}+ \frac{1}{2}\sqrt{\frac{2^{\gamma_c}}{| \mathcal {I}|}} .
\end{align}
The expectation of the last term can be evaluated as : 
\begin{align} 
&\Exp{ 1- \bar {P}_{I X' Y \hat{X}}( \mathcal {I}\times \mathcal {X}\times \mathcal {Y} \times \hat{\mathcal{X}})}{\mathcal {C}}\\ \notag
=&1-\mathbb {E}_ \mathcal {C}\bigg[\int_{\mathcal{X Y \hat{X}U}}\sum_i\frac{\mathbf{1}[u_i = u] \times \mathbf{1}\left[(u, x)\in \mathcal{T}_c(\gamma_c)\right]}{|\mathcal{I}|}.  \notag\\
&p_{X|U}(x|u)p_{Y|X}(y|x) p_{\hat{X}| U Y} ( \hat{x}| u,y) dxdyd\hat{x}du\bigg] \\
=&\prob{ (u,x) \notin \mathcal {T} _ \mathrm {c} (\gamma _ \mathrm {c})}{UX} .
\end{align}

Combining the results above provides the upper bound
\begin{align} 
\label{eq_apxe_bound}
&\varepsilon \le \mathbb{P}_{X\vec{U}\vec{Y} \hat {X}} \left [(u,y) \in \mathcal {T} _{ \mathrm {p}}(\gamma _{ \mathrm {p}})^{c} \cup (u,x) \in \mathcal {T} _{ \mathrm {c}}(\gamma _{ \mathrm {c}})^{c}  \right .\notag\\
&\left . \cup (x,\hat{x}) \in \mathcal {T} _{ \mathrm {d}}( {D})^{c} \cup (\vec{u}, \vec{y})\in \mathcal {T}_g (G)^{c}   \right ] + \frac {N}{2^{\gamma _{ \mathrm {p}}}| \mathcal {M}| } + \frac {1}{2}\sqrt {\frac {2^{\gamma _{ \mathrm {c}}}}{N}}. 
\end{align}

\section{Proof of theorem~\ref{thm_second_order_bound}}
\label{apx:second_order_coding_rate}
The proof is based on a Gaussian approximation using the following multi-dimensional Berry-Esséen theorem. 
\begin{theorem}[Multidimensional Berry-Esséen theorem\cite{Gotze_clt}]
Let $\vec{U}_1, \vec{U}_2, ... , \vec{U}_n$ be independent random vectors in $\mathbb{R}^k$ with zero mean. Let $\vec{S}_n = \frac{1}{\sqrt{n}}(\vec{U}_1+ ... + \vec{U}_n)$ and $\Cov{\vec{S}_n}{} = \mat{V} > \vec{0}$. Consider a Gaussian random vector $\vec{B}\sim \mathcal{N}(\vec{0}, \mat{V})$, then for all $n\in \mathbb{N}$, we have
\begin{align}
    \sup_{C\in \mathscr{C}_k} |\prob{C}{\vec{S}_n} - \prob{C}{\vec{B}_n}| \leq O\left(\frac{1}{\sqrt{n}}\right)
\end{align}
where $\mathscr{C}_k$ is the family of all convex Borel measurable subsets of $\mathbb{R}^k$.
\end{theorem}
    
As mentioned earlier, the components of the information-density-distortion-generalization error vector defined in \eqref{eq_idl} are not independent, since the generalization error $\Exp{\ell(\Tilde{X}, \hat{f}^{(n)}(\vec{Z},\Tilde{Y}))}{\tilde X \tilde Y}$ depends on the full sequence $\mathbf{Z}$, while the other components only depend on a random occurrence of $U, Y$. Therefore, let us consider the conditional information-density-distortion-generalization error vector rewritten as follow:
\begin{align}
    &\vec{j}_i(U_i, X_i, Y_i, \hat{X}_i|\vec{Z_{-i}})
    =\begin{bmatrix}\textstyle -\iota(u_i, y_i)\\ \iota(x_i, u_i)\\ d(x_i, \hat{x}_i)\\ \Exp{\ell(\Tilde{X}, \hat{f}^{(n)}(u_i, y_i, \Tilde{Y}))|\vec{Z_{-i}}}{\Tilde{X} \Tilde{Y}} \end{bmatrix}
\end{align}
where $\vec{Z_{-i}} = [\vec{u}^{i-1}, \vec{u}_{i+1}^n, \vec{y}^{i-1}, \vec{y}_{i+1}^n]$. Given that $U$ and $Y$ are independent random variables, let $\vec{Z}^\star = \vec{Z_{-i}} \sim P_{\vec{U}^{n-1}\vec{Y}^{n-1}}$. Its expectation $\vec{J}(\vec{Z}^\star)$ can be expressed as
\begin{align}
    \vec{J}(\vec{Z}^\star) &= \mathbb {E}[ \mathbf {j}_i(U_i,X_i,Y_i, \hat{X}_i|\vec{Z}_{-i})] =\begin{bmatrix} -I(U;Y)\\ I(U;X) \\ \Exp{d(X, \hat{X})}{X\hat{X}} \\ \Exp{\ell(\tilde{X}, \hat {f}^{(n)}(\vec{U}, \vec{Y}, \tilde{Y})|\vec{Z}_{-i}}{\Vec{Z}\tilde X \tilde Y} \end{bmatrix} .
\end{align} 

Let $\gamma_p = \log \frac{|\mathcal{I}_n|}{|\mathcal{M}_n|} + \log n$, where $\set{M}_n = \left\{1, \cdots, M_n \right\}$, and $\gamma_c = \log |\mathcal{I}_n| - \log n $, using~\eqref{eq_apxe_bound} allows us to show that there exists a code such that
\begin{align}
    &P_{e,n}(G, D) \leq \mathbb{P}_{X\vec{U}\vec{Y} \hat {X}} \left [(u,y) \in \mathcal {T} _{ \mathrm {p}}(\gamma _{ \mathrm {p}})^{c} \cup (u,x) \in \mathcal {T} _{ \mathrm {c}}(\gamma _{ \mathrm {c}})^{c}  \right .\notag\\
    &\ \left . \cup (x,\hat{x}) \in \mathcal {T} _{ \mathrm {d}}( {D})^{c} \cup (\vec{u}, \vec{y})\in \mathcal {T}_g (G)^{c}   \right ] +\frac{1}{n} + \frac{1}{2\sqrt{n}}\\
    &\leq \mathbb{E}_{\vec{Z^*}}\left[\mathbb{P}\left[\sum_i^n  \begin{bmatrix}-\iota(u_i,y_i)\\ \iota(x_i, u_i)\\ d(x_i, \hat{x}_i)\\ \Exp{\ell(\Tilde{X}, \hat{f}^{(n)}(u_i, y_i,\Tilde{Y}))|\vec{Z_{-i}}}{\Tilde{X}, \tilde{Y}} \end{bmatrix}\right. \right.
     \left. \left.\geq 
    \begin{bmatrix}\log \frac{|\mathcal{M}_n|}{|\mathcal{I}_n|}\\ \log |\mathcal{I}_n| \\ nD \\ nL \end{bmatrix} - \vec{\log n}  \right]\right]  + \frac{1}{n} + \frac{1}{2\sqrt{n}}\\
    &=\mathbb{E}_{\vec{Z^*}}\left[\mathbb{P}\left[\sum_i^n \left[\vec{j}_i - \vec{J}(\vec{Z}^\star) \right ] 
    \geq 
    \begin{bmatrix}\log \frac{|\mathcal{M}_n|}{|\mathcal{I}_n|}\\ \log |\mathcal{I}_n| \\ nD \\ nL \end{bmatrix} 
    - n\vec{J}(\vec{Z}^\star) - \vec{\log n}\right]\right]  + \frac{1}{n} + \frac{1}{2\sqrt{n}} .
\end{align}
Let 
\begin{align}
    \vec{\tilde b|Z^*} = \sqrt{n}\left [\begin{bmatrix}\frac{1}{n}\log \frac{|\mathcal{M}_n|}{|\mathcal{I}_n|}\\ \frac{1}{n} \log |\mathcal{I}_n| \\ D \\ L \end{bmatrix} - \vec{J}(\vec{Z}^\star) - \vec{\frac{2\log n}{n}} \right ] ,
\end{align}
where $\vec{\frac{\log n}{n}}$ denotes the vector $\frac{\log n}{n} \vec {1}_{4}$ with same size as vector $\vec{J}$.
We have
\begin{align}
    &1 - P_e(n, L, D)\\
    &\geq \prob{\prob{\frac{1}{\sqrt{n}}\sum_i^n \left[\vec{j}_i - \vec{J}(\vec{Z}^\star) \right ]
    \leq 
    \vec{\tilde b|Z^*} + \vec{\frac{\log n}{\sqrt{n}}}}{}}{\vec{Z^*}} 
     - \frac{1}{n} - \frac{1}{2\sqrt{n}}\\
    &=\Exp{\prob{\frac{1}{\sqrt{n}}\sum_i^n \left[\vec{j}_i - \vec{J}\textbf{} \right ]
    \leq 
    \vec{\tilde b|Z^*} + \vec{\frac{\log n}{\sqrt{n}}}}{\vec{J_i|Z^*}}}{\vec{Z^*}} 
     - \frac{1}{n} - \frac{1}{2\sqrt{n}}\\
    &\geq \prob{\prob{\vec{B|Z^*} \leq 
    \vec{\tilde b|Z^*} + \vec{\frac{\log n}{\sqrt{n}}}}{\vec{B|Z^*}}}{\vec{Z^*}}  - O\left(\frac{1}{\sqrt{n}}\right)\\
    &= \Exp{\prob{\vec{B|Z^*} \leq \vec{\tilde b |Z^*}}{\vec{B|Z^*}}}{\vec{Z^*}} + O\left(\vec{\frac{\log n}{\sqrt{n}}}\right)\\
    &=\Exp{\vec B\leq \vec {\tilde b}}{\vec{B}} + O\left(\vec{\frac{\log n}{\sqrt{n}}}\right)\\
    &\geq 1 - \epsilon ,
\end{align}
which indicates that 
\begin{align}
    \vec {\tilde b} &= \Exp{\vec{\tilde b}|\vec{Z^*}}{\vec{Z^*}} 
    = \sqrt{n}\left [\begin{bmatrix}\frac{1}{n}\log \frac{|\mathcal{M}_n|}{|\mathcal{I}_n|}\\ \frac{1}{n} \log |\mathcal{I}_n| \\ D \\ L \end{bmatrix} - \Exp{\vec{J}}{\vec{Z^*}} - \vec{\frac{2\log n}{n}} \right ]  \in  \mathscr{S}(\mat{V}, \epsilon)
\end{align}
where $\mathscr{S}(\mat{V}, \epsilon)$ is the dispersion region defined in \eqref{dispersion}. This completes the proof.

\section{Proof of Corollary \ref{cor_non_aymp}}
\label{apx:cor2}
    As shown in Theorem~\ref{thm_second_order_bound}, the bound for the second order coding rate is mainly affected by the dispersion region of $\vec{i}(X, \vec{U}, \vec{Y}, \hat{X})$, and especially by its covariance matrix. 
    In what follows, we aim to show that
    \begin{align}
        \label{covariance_ld}
        \text{Cov}\left (d(X, \hat{X}), G(\hat{f}^{(n)}, \vec{Z})\right) = 0 ,
    \end{align}
    which means that the distortion and the generalization error are uncorrelated. Here we provide the proof for both parametric regression with OLS estimator and kernel regression. Since $X, \hat{X}$ are i.i.d., without loss of generality, we consider $d(X, \hat{X}) = d(X_1, \hat{X_1})$.

    \subsection{Parametric regression }
    The covariance term \eqref{covariance_ld} can be expressed as: 
    \begin{align}
        &\text{Cov}\left (d(X_1, \hat{X}_1), G(\hat{f}^{(n)}, \vec{Z})\right)=
        &\Exp{G(\hat{f}^{(n)}, \vec{Z})d(X_1, \hat{X}_1)}{} - \Exp{G(\hat{f}^{(n)}, \vec{Z})}{}\Exp{d(X_1, \hat{X}_1)}{}
    \end{align}
    with
    \begin{align}
        &\Exp{G(\hat{f}^{(n)}, \vec{Z})}{}\Exp{d(X_1, \hat{X}_1)}{}
        = \bigg(\sigma^2 + \frac{\sigma^2 + \sigma_\Phi^2}{n}\Exp{\text{Tr}\left(\mat{\Tilde{\Sigma}}\mat{\Sigma}^{-1}\right)}{}\bigg) \Exp{d(X_1, \hat{X}_1)}{}
    \end{align}
    and 
    \begin{align}
        &\quad \Exp{G(\hat{f}^{(n)}, \vec{Z})d(X_1, \hat{X}_1)}{}\\ &= \sigma^2\Exp{d(X_1, \hat{X}_1)}{} 
        + \Exp{[\vec{\beta} - \vec{\hat{\beta}} ]^T\Exp{\vec{\tilde Y} \vec{\tilde Y}^T}{\tilde Y}[\vec{\beta} - \vec{\hat{\beta}} ] d(X_1, \hat{X}_1) }{}\\
        &= \sigma^2\Exp{d(X_1, \hat{X}_1)}{} + \frac{1}{n^2}\mathbb{E}_{Y}\left[\text{tr}\bigg(\tilde{\mat{\Sigma}}\mat{\Sigma}^{-1}\mat{Y}\right.
        \left.\Exp{(\vec{N} + \vec{\Phi})d(X_1, \hat{X}_1)(\vec{N} + \vec{\Phi})^T}{N\Phi}\mat{Y}^T\mat{\Sigma}^{-1}\bigg)\right]\\
        &= \sigma^2\Exp{d(X_1, \hat{X}_1)}{} + \frac{\sigma^2\sigma_\phi^2}{n^2}\Exp{\text{tr}\bigg(\tilde{\mat{\Sigma}}\mat{\Sigma}^{-1}\mat{Y}\mat{Y}^T\mat{\Sigma}^{-1}\bigg)}{}\label{expectation_ij}\\ 
        &=\sigma^2\Exp{d(X_1, \hat{X}_1)}{} + \frac{\sigma^2\sigma_\phi^2}{n}\Exp{\text{tr}\bigg(\tilde{\mat{\Sigma}}\mat{\Sigma}^{-1}\bigg)}{}
    \end{align}
    where \eqref{expectation_ij} follows from the fact that $\Exp{(\vec{N} + \vec{\Phi})d(X_1, \hat{X}_1)(\vec{N} + \vec{\Phi})^T}{N\Phi}$ remains the same for all $i \in \llbracket 1, n\rrbracket$. 
    Using the fact that $\sigma^2\sigma_\phi^2 = (\sigma^2 + \sigma_\phi^2)\Exp{d(X_1, \hat{X}_1)}{}$ completes the proof for the parametric case.

    \subsection{Kernel regression}
    For kernel regression, we express: 
    \begin{align}
        &\quad \Exp{G(\hat{f}^{(n)}, \vec{Z})}{\Tilde{X}\Tilde{Y} \vec{Z}}\Exp{d(X_1, \hat{X}_1)}{\vec{Z}}\\
        &= \left(\sigma^2 + \Exp{{f}^2\left(\tilde Y\right)}{\tilde Y }+ \Exp{\hat{f}^{(n)^2}(\tilde Y)}{\tilde Y \vec{Z}} \right.
        \notag\\
        &\quad\left.- 2\Exp{{f}\left(\tilde y\right)\Exp{\hat{f}^{(n)}\left(\tilde y\right) }{\vec{Z}} \left| \Tilde{Y} = \Tilde{y} \right. }{\tilde Y }\right) \Exp{d(X_1, \hat{X}_1)}{\vec{Z}}, \notag
    \end{align}
    and 
    \begin{align}
        &\quad \Exp{G(\hat{f}^{(n)}, \vec{Z})d(X_1, \hat{X}_1)}{\Tilde{X}\Tilde{Y} \vec{Z}}\\\notag \label{eq_apx_g_gd}
        &=\sigma^2\Exp{d(X_1, \hat{X}_1)}{\vec{Z}}+ \Exp{{f}^2(\tilde Y)}{\tilde Y }\Exp{d(X_1, \hat{X}_1)}{\vec{Z}}\\
        &\quad + \Exp{\Exp{\hat{f}^{(n)^2}(\tilde y)d(X_1, \hat{X}_1)}{\vec{Z}}\left| \Tilde{Y} = \Tilde{y} \right.}{\tilde Y }\\
        &\quad- 2\Exp{{f}(\tilde y)\Exp{\hat{f}^{(n)}(\tilde y)d(X_1, \hat{X}_1)}{\vec{Z}} \left| \Tilde{Y} = \Tilde{y} \right.}{\tilde Y }
    \end{align}
    For the last term, 
    we have
    \begin{align}
        &\quad \Exp{\hat{f}^{(n)}(\tilde y)d(X_1, \hat{X}_1)}{\vec{Z}}\\ \label{eq_apx_g_1}
        &=f(\Tilde{y})\Exp{d(X_1, \hat{X}_1)}{\vec{Z}} + \Exp{\frac{\hat{m}_1(\tilde y)}{\hat{p}(\tilde y)}}{\vec{Y}}\Exp{d(X_1, \hat{X}_1)}{\vec{N}\vec{\Phi}}
        + \Exp{\frac{\hat{m}_2(\tilde y)d(X_1, \hat{X}_1)}{\hat{p}_Y(\tilde y)}}{\vec{Y}\vec{N}\vec{\Phi}}\\
        &= f(\Tilde{y})\Exp{d(X_1, \hat{X}_1)}{\vec{Z}} + \Exp{\frac{\hat{m}_1(\tilde y)}{\hat{p}_Y(\tilde y)}}{\vec{Y}}\Exp{d(X_1, \hat{X}_1)}{\vec{N}\vec{\Phi}} \notag\\
        &\quad+ \sum_{i=1}^n \Exp{\frac{K(\frac{\Tilde{y}-Y_i}{h})}{nh\hat{p}(\tilde y)}}{\vec{Y}} \Exp{(N_i+\Phi_i))d(X_1, \hat{X}_1)}{\vec{N}\vec{\Phi}}\\
        &=f(\Tilde{y})\Exp{d(X_1, \hat{X}_1)}{\vec{Z}} + \Exp{\frac{\hat{m}_1(\tilde y)}{\hat{p}_Y(\tilde y)}}{\vec{Y}}\Exp{d(X_1, \hat{X}_1)}{\vec{N}\vec{\Phi}} \label{pf_apx_g_2}\\ 
        &=\Exp{\hat{f}^{(n)}(\tilde y)}{\vec{Z}}\Exp{d(X_1, \hat{X}_1)}{\vec{Z}} ,
    \end{align}
    where \eqref{eq_apx_g_1} comes from the fact that $d(X_1, \hat{X}_1)$ is independent from $ \hat{p}_Y(\Tilde{y})$ since $N$ and $\Phi$ are independent from $Y$. In addition,~\eqref{pf_apx_g_2} is because for all $i \in \llbracket 1, n\rrbracket$, $\Exp{(N_i+\Phi_i))d(X_1, \hat{X}_1)}{\vec{N}\vec{\Phi}} = 0$.

    Then for the third term \eqref{eq_apx_g_gd}, we have
    \begin{align}
        &\quad\Exp{\hat{f}^{(n)^2}(\tilde y)d(X_1, \hat{X}_1)}{\vec{Z}} \\ \label{eq_apxg_1}
        &=\Exp{f^2(\Tilde{y}) + \left(\frac{\hat{m}_1(\tilde y)}{\hat{p}_Y(\tilde y)}\right)^2 + 2\frac{f(\Tilde{y})\hat{m}_1(\Tilde{y})}{\hat{p}_Y(\tilde y)} }{} \Exp{d(X_1, \hat{X}_1)}{} \\
        &\quad+ \Exp{\left( 2\frac{f(\Tilde{y})\hat{m}_2(\Tilde{y})}{\hat{p}_Y(\tilde y)} + 2\frac{\hat{m}_1(\Tilde{y})\hat{m}_2(\Tilde{y})}{\hat{p}_Y(\tilde y)}\right)d(X_1, \hat{X}_1)}{}\\ \label{eq_apxg_2}
        &\quad+ \Exp{ \left(\frac{\hat{m}_2(\tilde y)}{\hat{p}_Y(\tilde y)}\right)^2 d(X_1, \hat{X}_1)}{}
    \end{align}
    where \eqref{eq_apxg_1} is obtained from the same arguments as for \eqref{eq_apx_g_1}, and the second part equals to zero because of equation \eqref{pf_apx_g_2}. The last part \eqref{eq_apxg_2} can be developped as 
    \begin{align}
        &\quad \Exp{ \left(\frac{\hat{m}_2(\tilde y)}{\hat{p}_Y(\tilde y)}\right)^2 d(X_1, \hat{X}_1)}{\vec{Z}}\\
        &=\frac{1}{n^2h^2}\Exp{\sum_{i=1}^n\left(\frac{K(\frac{\Tilde{y}-Y_i}{h})(N_i + \Phi_i)}{\hat{p}_Y(\tilde y)}\right)^2d(X_1, \hat{X}_1)}{\vec{Y}\vec{N}\vec{\Phi}}\\
        &=\frac{1}{n^2h^2}\Exp{\sum_{i=1}^n\left(\frac{K(\frac{\Tilde{y}-Y_i}{h})}{\hat{p}_Y(\tilde y)}\right)^2}{\Vec{Y}} 
        \Exp{(N_i + \Phi_i)^2\left((\alpha-1)N_j + \alpha \Phi_j\right)^2}{\vec{N}\vec{\Phi}}\\ \label{pf_apx_g_3}
        &=\frac{\sigma^2 + \sigma_\Phi^2}{n^2h^2}\Exp{\sum_{i=1}^n\left(\frac{K(\frac{\Tilde{y}-Y_i}{h})}{\hat{p}_Y(\tilde y)}\right)^2}{\Vec{Y}}\Exp{d(X_1, \hat{X}_1)}{\vec{Z}}\\
        &=\Exp{ \left(\frac{\hat{m}_2(\tilde y)}{\hat{p}_Y(\tilde y)}\right)^2 }{\vec{Z}}\Exp{d(X_1, \hat{X}_1)}{\vec{Z}}
    \end{align}
    with \eqref{pf_apx_g_3} is obtained from the same arguments as for~\eqref{expectation_ij}.

   This gives 
   \begin{align}
       &\Exp{{f}(\tilde y)\Exp{\hat{f}^{(n)}(\tilde y)d(X_1, \hat{X}_1)\left| \Tilde{Y} = \Tilde{y} \right.}{\vec{Z}} }{\tilde Y }\notag\\
       =& \Exp{{f}(\tilde y)\hat{f}^{(n)}(\tilde y) }{\tilde Y \vec{Z}}\Exp{d(X_1, \hat{X}_1)}{\vec{Z}},
   \end{align}
   therefore 
   \begin{align}
       &\Exp{G(\hat{f}^{(n)}, \vec{Z})d(X, \hat{X})}{\Tilde{X}\Tilde{Y} \vec{Z}} \notag\\
       =& \Exp{G(\hat{f}^{(n)}, \vec{Z})}{\Tilde{X}\Tilde{Y} \vec{Z}}\Exp{d(X,\hat{X})}{\vec{Z}},
   \end{align}
   and $\text{Cov}\left (d(X, \hat{X}), G(\hat{f}^{(n)}, \vec{Z})\right) = 0$.
    
   Denote $R_b(n,D,G,\varepsilon)$ as the infimum introduced by Theorem \ref{thm_second_order_bound} for the rate-distortion-generalization error function, $R_b(n, D, \varepsilon)$ and $R_b(n, G, \varepsilon)$ for the rate-distortion function and rate-generalization error function, respectively. Consider the vector $\vec{B} = [B_1, B_2, B_3, B_4]^T$ defined in \eqref{dispersion}, by the achivability of $R_b(n, D, \varepsilon)$ and $R_b(n, G, \varepsilon)$, we have :
   \begin{align}
       &\prob{B_1\leq b_1, B_2 \leq b_2, B_3\leq D}{B_1B_2B_3} = 1-\varepsilon, \\
       &\prob{B_1\leq b_1', B_2 \leq b_2', B_4\leq G}{B_1B_2B_4} = 1-\varepsilon
   \end{align}
   with $R_b(n, D, \varepsilon) = I(X; U) - I(U; Y) + b_1 + b_2 + O\left(\frac{\log n}{n}\right)$ and $R_b(n, G, \varepsilon) = I(X; U) - I(U; Y) + b_1' + b_2' + O\left(\frac{\log n}{n}\right)$. 
   
   Consider firstly $R_b(n, D, \varepsilon) \geq R_b(n, G, \varepsilon)$ and $R_b = \max\{R_b(n, D, \varepsilon), R_b(n, G, \varepsilon)\}$, we have
   \begin{align}
       &\prob{B_1\leq b_1, B_2 \leq b_2, B_3\leq D, B_4\leq G}{\vec{B}}\\
       =&\prob{B_1\leq b_1, B_2 \leq b_2}{}\prob{B_3\leq D|B_1\leq b_1, B_2 \leq b_2}{}
       \prob{B_4\leq G|B_1\leq b_1, B_2 \leq b_2}{}\label{eq_apxg_idp}\\
       =&\prob{B_1\leq b_1, B_2 \leq b_2, B_3\leq D}{}
       \prob{B_4\leq G|B_1\leq b_1, B_2 \leq b_2}{}\\
       =&(1 - \varepsilon) \prob{B_4\leq G|B_1\leq b_1, B_2 \leq b_2}{}\\
       <&1 -\varepsilon \label{eq_apxg_cdfg},
   \end{align}
   where \eqref{eq_apxg_idp} follows the fact that uncorrelation of Gaussian variables indicates independence, \eqref{eq_apxg_cdfg} is because the cumulative density function of Gaussian source is smaller than 1.

   The same analysis applies for the case $R_b(n, G, \varepsilon) \le R_b(n, D, \varepsilon)$. It implies that in order to ensure the same excess probability, a higher rate $R_b(n, D, G, \varepsilon)$ is necessary by the approximation of Berry-Esséen Theorem, that is 
   \begin{align}
       R_b(n, D, G, \varepsilon) > \max\{R_b(n, D, \varepsilon), R_b(n, G, \varepsilon)\}
   \end{align}
   
   This completes the proof.

\bibliographystyle{IEEEtran}
\bibliography{references}

\end{document}